\documentclass[pra,amsmath,amssymb,preprint]{revtex4-2}

\usepackage{graphicx}
\usepackage{dcolumn}
\usepackage{bm}
\usepackage[mathlines]{lineno}
\usepackage{tikz}
\usepackage[utf8]{inputenc}
\usepackage[T1]{fontenc}
\usepackage{mathptmx}
\usepackage{etoolbox}
\usepackage{upgreek,bm}
\usepackage{xcolor}
\usepackage{ulem,xpatch}
\usepackage{epstopdf}

\makeatletter
\def\@email#1#2{%
 \endgroup
 \patchcmd{\titleblock@produce}
  {\frontmatter@RRAPformat}
  {\frontmatter@RRAPformat{\produce@RRAP{*#1\href{mailto:#2}{#2}}}\frontmatter@RRAPformat}
  {}{}
}%
\makeatother

\begin{document}

\title{Third order nonlinear correlation of the electromagnetic vacuum at near-infrared frequencies}

\author{Francesca Fabiana Settembrini}
 \email{fsettemb@phys.ethz.ch}
\author{Alexa Herter}%
\author{Jér\^{o}me Faist}
 \email{jfaist@phys.ethz.ch}
\affiliation{ETH Z\"urich, Institute of Quantum Electronics, Auguste-Piccard-Hof 1, 8093 Z\"urich, Switzerland}

\date{\today}

\begin{abstract}
 
In recent years, electro-optic sampling, which is based on Pockel’s effect between an electromagnetic mode and a copropagating, phase-matched ultrashort probe, has been largely used for the investigation of broadband quantum states of light, especially in the mid-infrared and terahertz frequency range. The use of two mutually delayed femtosecond pulses at near-infrared frequencies allows the measurement of quantum electromagnetic radiation in different space-time points. Their correlation allows therefore direct access to the spectral content of a broadband quantum state at THz frequencies after Fourier transformation. 
In this work, we will prove experimentally and theoretically that when using strongly focused coherent ultrashort probes, the electro-optic sampling technique can be affected by the presence of a third-order nonlinear mixing of the probes' electric field at near-infrared frequencies. Moreover, we will show that these third-order nonlinear phenomena can also influence correlation measurements of the quantum electromagnetic radiation. We will prove that the four-wave mixing of the coherent probes' electric field with their own electromagnetic vacuum at near-infrared frequencies results in the generation of a higher-order nonlinear correlation term. The latter will be characterized experimentally, proving its local nature requiring the physical overlap of the two probes. The parameters regime where higher order nonlinear correlation results predominant with respect to electro-optic correlation of terahertz radiation is provided.

\end{abstract}

\maketitle
\newpage

\section{Introduction}

Nonlinear quantum optics \cite{Chang2014} has been of paramount importance for recent technological developments in a multitude of research fields, from optical quantum communication \cite{Kimble2008}, quantum computing \cite{Knill2001} to quantum spectroscopy\cite{Schlawin2017} and quantum metrology \cite{Giovannetti2011}. In particular, the optimization of nonlinear photon-photon interaction has been crucial for the implementation of new platforms for quantum logic integrated on-chip \cite{Miller2010,Uppu2020,Luo2019} and of building blocks for efficient entanglement and squeezed radiation generation, which is fundamental for improved performances in quantum sensing \cite{Pirandola2018}.
 
Nonlinear interaction of photons at different frequencies has been particularly significant for the metrological study of fundamental states of electromagnetic radiation, especially in the mid-infrared (MIR) and terahertz (THz) frequency range. In alternative to commonly used heterodyne detection, an established measurement scheme compatible with the investigation of the most fundamental broadband states of electromagnetic radiation, has been developed through electro-optic sampling (EOS) \cite{Gallot1999}. This measurement technique exploits the Pockels effect in a material with $\chi^{(2)}$ nonlinearity to map the amplitude of the investigated electromagnetic field on the polarization state of a phase-matched femtosecond pulse. The use of ultrashort laser pulses insures the subcycle resolution of the electromagnetic radiation under study both in space and in time and the nonlinear properties of the material determine the large detection bandwidth of the technique.

Experimental implementations of electro-optic detection have led to the first measurements of the statistical properties of the electromagnetic vacuum in the MIR frequency range \cite{Riek2015,Riek2017}, through a parametric study of the detection system's shot noise. 
These results have led to further theoretical proposals in the field of quantum metrology for the exploration of higher order noise distributions of the electromagnetic vacuum \cite{Kizmann2019,Guedes2019,Sulzer2020,Virally2021}.

In previous works, we have presented a further development of the electro-optic field detection technique involving the use of two probing pulses, which enables the measurement of electromagnetic radiation in distinct space-time points. This opened up the possibility of investigating both second- \cite{Benea-Chelmus2016} and first-order coherence on an electromagnetic quantum state of radiation, which provides access to its spectral content after a Fourier transformation. We have experimentally proven that the developed electro-optic field correlation measurement scheme allows access to the spatial and temporal coherence of the electromagnetic quantum vacuum in the THz frequency range \cite{Benea-Chelmus2019}. In particular, our latest results \cite{Settembrini2022} have provided the first experimental proof of a fundamental hypothesis in quantum electrodynamics, which claims the quantum vacuum to be correlated outside the relativistic light cone \cite{Valentini1991}.

In order to improve the accuracy of quantum metrology measurements based on electro-optic sampling, a strong confinement of the probing radiation in both space and time is needed. However, the presence of a strong local probing electric field will influence the measured electromagnetic state through quantum back-action \cite{Guedes2022} as well as lead to the appearance of competing higher-order nonlinear phenomena. In fact, in a recent theoretical work \cite{Gundogdu2022}, the higher order nonlinear mixing of the electromagnetic radiation studied with the local probing laser field has been proposed in combination with homodyne detection as an alternative method for quantum noise distribution measurements with efficient background suppression.

In this work, we will present the experimental results of electro-optic electric field correlation measurement on a thermally populated electromagnetic state, obtained using highly confined probing laser beams with a strong spatial overlap. Our results present a significant deviation from the expected electro-optic field correlation induced by thermal radiation at THz frequencies both in time and frequency domain. We will prove that the observed results can be ascribed to the effect of higher-order nonlinear correlation of the probing pulses' electric field with the electromagnetic vacuum at their own near-infrared frequency. In order to validate our hypothesis, the dependence of the detected field correlation measurement on the experimental parameters will be investigated and the parameters range in which third-order nonlinear phenomena become predominant with respect to electro-optic detection will be defined.

The paper is organized as follows. In Section II, we investigate both theoretically and experimentally the effect of the third-order non-linearity onto the balanced detection scheme by modulating one beam and detecting its influence on the copropagating one. In the third section, we investigate how the third-order non-linearity influences the correlation of their fluctuations and show that the nonlinear correlation signal still arises from vacuum fluctuations - albeit at near-infrared frequencies. We also show that the signal arises only from the physical overlap of the two probing beams, a feature that is exploited in the non-local measurements of vacuum fluctuations.   

\section{Third order nonlinear balanced detection} \label{Sec:ThirdOrderClassical}

\begin{figure*}
    \centering
    \includegraphics{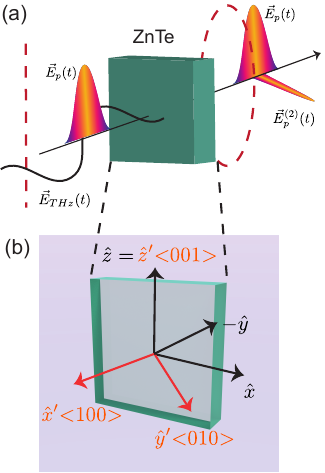}
    \caption{\textit{Balanced ellipsometry detection}. a)  Electro-optic detection in ZnTe is determined by the change in polarization (red dashed line) of a femtosecond probe pulse $\vec{E}_\mathrm{p}(t)$ due to its second order nonlinear interaction with a phase matched THz electric field  $\vec{E}_\mathrm{THz}(t)$ (solid black line). b) Relative orientation of the laboratory axis (in black) and ZnTe crystallographic axis (in orange) with respect to the facet of the crystal.}
    \label{Fig:ClassicKerrSetup}
\end{figure*}

In our experimental implementation of electro-optic field coherence detection, the nonlinear medium chosen is a $\langle 110\rangle$-cut zinc telluride (ZnTe) crystal. The working principle of electro-optic sampling implemented with the ZnTe crystal is shown schematically in Fig.~\ref{Fig:ClassicKerrSetup} (a). The interaction of an electromagnetic mode $\vec{E}_\mathrm{THz}(t)$ with a phase-matched femtosecond probing pulse, described by the electric field $\vec{E}_\mathrm{p}(t)$, leads to the creation of a second order nonlinear polarization $\vec{P}^{(2)}(t)$. The resulting additional electric field component $\vec{E}^{(2)}(t)$ at the frequency of the probe is oriented along the perpendicular direction with respect to the original pulse polarization. The subsequent change in polarization from linear to elliptical can be then measured via a homodyne detection system based on balanced ellipsometry \cite{vanderValk2004}. The combination of a quarter-wave plate and a polarizing beam-splitter mixes the NIR $z$-component of the probe electric field $\vec{E}_\mathrm{p}(t)$ (local oscillator) with NIR $x$-component $\vec{E}^{(2)}(t)$ generated in the nonlinear process. The two beams separated by the polarizing beam splitter both contain the mixing of local oscillator and signal. As a consequence, the subtraction of the beams measured at the two photodiodes of the balanced detector removes the local oscillator's intensity, while a signal proportional to the amplitude of the NIR $\vec{E}^{(2)}(t)$ field is measured. 

In the frequency domain, the electro-optic detection process relies on sum and difference frequency generation \cite{Gallot1999}. The nonlinear interaction of a single mode $\omega$ within the broadband femtosecond probe spectrum with the THz electromagnetic mode at a much lower frequency $\Omega$ leads to the creation of modes $\omega \pm \Omega$ still lying within the spectral bandwidth of the ultrashort pulse, therefore allowing their direct detection.
 
The crystallographic axes of ZnTe together with the laboratory reference frame are presented in Fig.~\ref{Fig:ClassicKerrSetup} (b) in orange and black respectively. In order to achieve maximum sensitivity and avoid the creation of additional radiation at THz frequency via optical rectification~\cite{Planken2001}, in all of the presented experimental work the polarization of the probing pulses has been oriented along the laboratory $z$-axis (as indicated also in Fig.~\ref{Fig:ClassicKerrSetup} (a)). Due to the zinc blende symmetry of ZnTe and the chosen polarization of the probes, the electro-optic detection results in a measurement sensitive exclusively to THz radiation polarized along the $\hat{x}$ axis in the laboratory reference frame.

The use of a balanced ellipsometry detection implicitly allows the nonlinear detection scheme to be susceptible to any higher-order nonlinear phenomena which could lead to a probe polarization change with equal crystallographic symmetry and resulting frequency within the probe pulse bandwidth as electro-optic sampling.

Significant third-order non-linearities have been observed in ZnTe crystals using tightly focused pulsed near-infrared radiation\cite{Chen2009}. Using a pump-probe detection scheme involving two mutually delayed ultrashort pulses, it has been experimentally shown that a four-wave mixing process between a high-intensity pump and a weaker probe electric fields can lead to the generation of a signal detectable via balanced ellipsometry. For probing pulses with a temporal extent in the order of $100~$fs, Gaussian beam waist smaller than $100~\upmu$m and average optical powers in the order of tens of mW, third-order nonlinear interaction has been demonstrated to produce a probe polarization change comparable in magnitude with the one induced via electro-optic detection of THz radiation~\cite{Caumes2002,Zhen2008}. The two contributions however present significantly different features in the temporal domain. In particular, the balanced signal caused by third-order nonlinearities has been theoretically demonstrated to be directly proportional to the intensity autocorrelation trace of the two pulses in the time domain. 

\subsection{Semi-classical description of third order nonlinear balanced signal}\label{Chapt:Semi-ClassKerrTheory}

In order to predict theoretically the influence that the third-order nonlinear interaction bears on balanced detection in our experimental implementation of electro-optic sampling, we follow the derivations presented in Ref.~\cite{Moskalenko2015,Caumes2002}. We assume the use of two ultrashort probing pulses, mutually delayed by a time $\tau$ and both polarized along the $\hat{z}$ direction of the laboratory reference axis. We will furthermore assume the presence of a non-zero probe electric field component along the $\hat{x}$ axis for the initial pulse ($t$ pulse): $\vec{E}_t(t)=(E_{t,x}(t),0,E_{t,z}(t)),\vec{E}_{\tau}(t+\tau)=(0,0,E_{\tau,z}(t+\tau))$. For simplicity, we will focus our analysis only on the dependence of the nonlinear signal detected by the time-delayed pulse ($t+\tau$ pulse) due to its third-order nonlinear interaction with its non-delayed version ($t$ pulse). 

According to the crystallographic symmetry of ZnTe \cite{Caumes2002}, the higher-order nonlinear interaction between the two ultrashort probes will result in the creation of the following third-order polarization terms (for further details see  Sec. I of the Suppl. Material):
\begin{subequations}
\begin{equation}
\postdisplaypenalty=0 
P^{(3)}_x(t+\tau)=2\epsilon_0\chi_{44}E_{t,z}(t)E_{t,x}(t)E_{\tau,z}(t+\tau),
\label{eq:TONPol}   
\end{equation}
\begin{equation}
 P^{(3)}_z(t+\tau)=\epsilon_0\chi_{11}E_{t,z}(t)E_{t,z}(t)E_{\tau,z}(t+\tau).
 \label{eq:TONPolz} 
\end{equation}
\end{subequations}

Here, $\epsilon_0$ represents the vacuum permittivity and the terms $\chi_{11}, \chi_{44}$ are the only non-vanishing components of the third-order nonlinear tensor in ZnTe. Their values correspond to $\chi_{11}= 3 \times 10^{-19} \frac{\mathrm{m}^2}{\mathrm{V}^2}$ and $\chi_{44}=1.5\times 10^{-19} \frac{\mathrm{m}^2}{\mathrm{V}^2}$, respectively~\cite{Caumes2002}. Eq.~(\ref{eq:TONPol}) clearly points to the equivalence of the crystallographic symmetry of third-order nonlinear interaction and electro-optic detection. It is important to underline as well the presence of an additional polarization component oriented along the $\hat{z}$ laboratory axis, presented in Eq.~(\ref{eq:TONPolz}). Due to the balanced detection scheme, the latter will not influence the result of the final ellipsometry measurement but will only be responsible for an intensity modulation of the probe and will be therefore disregarded in the following analysis. 

Following the derivations' steps presented in Ref.~\cite{Moskalenko2015} in the case of classical fields, the nonlinear signal detected by the $t+\tau$ probing pulse $S^{(3)}(\tau)$ will assume the form:
\begin{equation}
S^{(3)}(\tau)= \frac{1}{2} c n \epsilon_0 \int_{0}^{\infty} d \omega \frac{\eta(\omega)}{\hbar \omega} \int d^2 \Vec{r_{\perp}}|E_{\tau,z}(\zeta)|^2 \left[ i\;\frac{ E^{(3)}_{\tau,x}(\zeta)}{E_{\tau,z}(\zeta)}+\mathrm{h.c.}\right].
\label{eq:KerrEOSignalGen}
\end{equation}
In this expression, $c$ indicates the speed of light, $n$ the refractive index of ZnTe crystal at NIR frequencies, which are indicated as $\omega$. The function $\eta(\omega)$ represents the quantum responsivity of the balanced detector in the NIR frequency region. The coordinates $\vec{r}_{\perp} = (x,0,z)$ indicate the spatial position on the transverse plane with respect to the propagation direction of the ultrashort probing pulses $\hat{y}$ in the laboratory reference frame. The set of coordinates $\zeta$ is defined as $\zeta=\{ (x,z),l,\omega\}$, where $l$ is the length of the crystal, and represents the position of the two probes after propagation through the nonlinear medium. The quantity $E^{(3)}_{\tau,x}(\zeta)$ represents the perpendicular electric field component acquired by the time-delayed $t+\tau$ probe after propagation in the detection crystal and generated by the nonlinear polarization term described in Eq.~(\ref{eq:TONPol}). Assuming for the two interacting ultrashort probes a Gaussian distribution both in space and in the frequency domain (see Suppl. Material Sec. I for details), the expression in Eq.~(\ref{eq:KerrEOSignalGen}) can be simplified as follows:
\begin{equation}
   S^{(3)}(\tau) =\frac{\chi_{44} \omega_\mathrm{p} N_{\tau}E_{t,x}E_{t,z}}{n c w_0^2} \int^{+\infty}_{-\infty}\int^{+\infty}_{-\infty}d\omega^{\prime}~ d\omega^{\prime\prime}~ \Gamma(\omega^{\prime}, \omega^{\prime\prime}) A_\mathrm{overlap}.
   \label{eq:KerrClassicFinal}
\end{equation}
Here, $\omega_\mathrm{p}$ represents the central frequency of the ultrashort probing pulse and $w_0$ its Gaussian waist. The amplitude at focus of the electric fields' components of the non-time delayed probe along the $\hat{x},\hat{z}$ direction is indicated as $E_{t,x},E_{t,z}$ respectively. The quantity $N_{\tau}$ represents the total number of photons in the temporal delayed pulse impinging on the balanced detector. The function $A_\mathrm{overlap}$ describes the spatial overlap of the two ultrashort probe pulses propagating along the nonlinear material (see Suppl. Material Sec.I Eq.~(19)). The function $\Gamma(\omega^{\prime}, \omega^{\prime\prime})$ indicates the convolution of the spectral distributions of the electric fields involved in the nonlinear mixing and effectively represents the spectral intensity autocorrelation function of the two pulses. 

As predicted in Ref.~\cite{Caumes2002}, Eq.~(\ref{eq:KerrClassicFinal}) indicates the possible presence of a higher order nonlinear balanced signal $S^{(3)}(\tau)$ due to the interaction of two ultrashort probing pulses polarized along the $\hat{z}$ axis in ZnTe. As it is characteristic for third-order nonlinear phenomena, the amplitude of the signal generated results directly proportional to the product of the two electric field components $E_{t,x}E_{t,z}$ (and therefore to the overall intensity) of the t femtosecond pulse. This dependence also clarifies the direct connection between the arising of the higher order nonlinear balanced signal $S^{(3)}(\tau)$ and the presence of a non-vanishing component of the t probing pulse along the $\hat{x}$ direction. Moreover, the mutual interaction of the two laser probes leads to the generation of a signal $S^{(3)}(\tau)$ which represents in the time domain their intensity autocorrelation function, as derived in Eq.~(16) of Sec. I in the Suppl. Material.  

\subsection{Experimental study of third order nonlinear balanced signal}\label{Chpt:Semi-ClassicalKerrExp}

\begin{figure}
    \centering
    \includegraphics{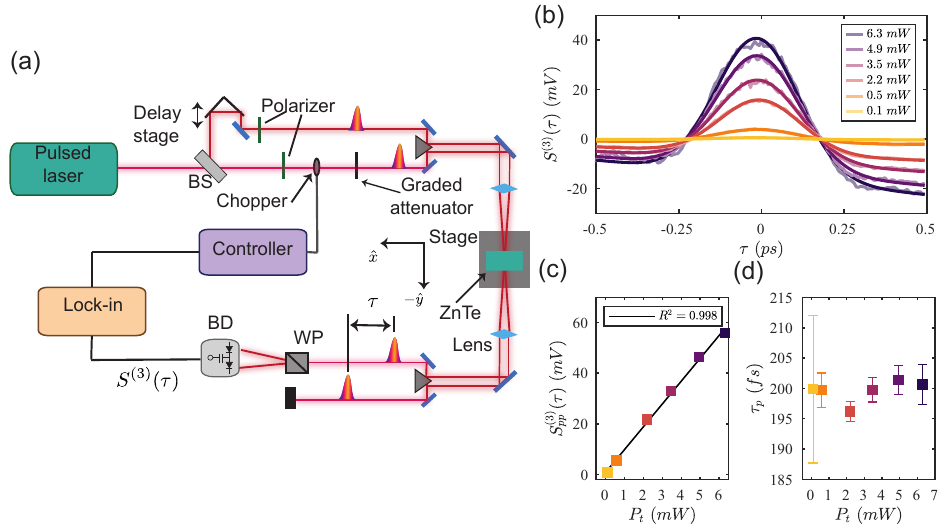}
    \caption{\textit{Third order balanced coherent detection}. a) Experimental setup used for coherent nonlinear signal detection via balanced ellipsometry. Both $t$ and $t+\tau$ probes are polarized along the $<001>$ axis of the ZnTe crystal, which coincides with the $z$-axis of the laboratory reference frame. The acquisition of the signal has been performed with a lock-in at the optical chopping frequency of $f = 600 ~$Hz. BD = balanced detectors, WP = Wollaston prism, WVP = quarter waveplate. (b) Experimental third-order balanced signal $S^{(3)}(\tau)$ recorded for different probing power of the interacting pulse $P_t$ (faded line) and their respective Gaussian fitting functions (solid lines). The experimental timetraces have been recorded with a temporal resolution of $33~$fs and an integration time of $2~$s per point. (c) The peak-to-peak amplitude of the experimental signal $S^{(3)}_\mathrm{pp}(\tau)$ presents a linear dependence with the intensity of the copropagating pulse $P_t$. The uncertainty on the experimental measurement is derived from the uncertainty of the fitting parameters. (d) The extracted temporal extent of the two interfering femtosecond pulses $\tau_\mathrm{p}$ is retrieved from the Gaussian fitting function $g(\tau)$. The uncertainty on the reported value represents the $2\sigma$ confidence interval and it has been obtained from the uncertainty of the fitting parameters.}
    \label{Fig:ClassicKerrResults}
\end{figure}

In order to verify the presence of the third order induced balanced detection signal $S^{(3)}(\tau)$ predicted in Eq.~(\ref{eq:KerrClassicFinal}) we have implemented the experimental setup sketched in Fig.~\ref{Fig:ClassicKerrResults} (a). The setup is similar to the one reported in Ref.~\cite{Zhen2008}. The ultrashort pulsed radiation generated by a Ti:Sapphire laser at a wavelength of $800~$nm is divided into two equal optical paths. A delay stage on one of these paths allows an adjustable temporal delay $\tau$  between the two identical femtosecond pulses. They are subsequently collected by a system of short focal-length lenses and focused with a Gaussian beam waist of $w_0 = 10~\upmu$m in the 1 mm long ZnTe crystal placed at the center of the lens system. The influence of the third-order nonlinear interaction between the femtosecond probes can be investigated by acquiring the balanced ellipsometry signal registered by the time delayed $t+\tau$ probe $S^{(3)}(\tau)$ via a lock-in acquisition system. No external source at THz frequencies is present and the generation of THz electromagnetic radiation via optical rectification is inhibited by the choice of the probes polarization direction along the $\hat{z}$ laboratory reference axis. The two femtosecond probes present as well a residual electric field component along the $\hat{x}$ laboratory axis, due to the non-perfect extinction ratio of the polarizers. The polarization components of the two probes present values of $P_{t,z} = 6.3~$mW, $P_{\tau,z} = 7.2~$mW and $P_{t,x} = 170~\upmu$W, $P_{\tau,x} = 120~\upmu$W at the detection crystal facet along the $\hat{z}$, $\hat{x}$ axis of the laboratory reference frame respectively.

The experimental balanced ellipsometry signal $S^{(3)}(\tau)$ recorded as a function of the pulses temporal delay $\tau$ is presented for different powers of the copropagating t probing pulse $P_t$ in Fig.~\ref{Fig:ClassicKerrResults} (b). All experimental results have been collected maintaining the optical power of the detection $t+\tau$ pulse constant. As it can be clearly observed in Fig.~\ref{Fig:ClassicKerrResults} (b), the amplitude of the nonlinear signal $S^{(3)}(\tau)$ appears to decrease significantly as a function of the t probe power $P_t$. In order to obtain a more accurate estimation of the $S^{(3)(\tau)}$ amplitude dependence, the experimental data in Fig.~\ref{Fig:ClassicKerrResults} (b) have been fitted with a Gaussian function of the form $g(\tau) = c+a\tau+b \exp\left(-\frac{4\ln(2)(\tau-d)^2}{\gamma^2}\right)$. As clearly shown in Fig.~\ref{Fig:ClassicKerrResults} (c), the peak-to-peak amplitude of the fitted signal $S^{(3)}_\mathrm{pp}(\tau)$ with respect to the baseline $a\tau$ presents a direct proportionality to the intensity of the interacting probing pulse $P_t$, decreasing from a value of $56~$mV for $P_t = 6.3~$mW to a value of $0.76~$mV in the case of $P_t = 0.1~$mW. This experimental result is in good accord with the intensity dependence of a signal generated via third-order nonlinear interaction, as reported in Eq.~(\ref{eq:KerrClassicFinal}).

As proposed in Ref.~\cite{Caumes2002} and derived in Eq.~(\ref{eq:KerrClassicFinal}), the experimental data reported in Fig.~\ref{Fig:ClassicKerrResults} (b) should also correspond in the temporal domain to the intensity autocorrelation of the two interacting pulses $\langle I(t)I(t+\tau)\rangle$. From the fitting parameter $\gamma$, the temporal extent of the two ultrashort NIR probes can be extracted through the relation $\tau_\mathrm{p} = 0.7\gamma$ ~\cite{siegman1986}. The results obtained from the experimental data reported in Fig.~\ref{Fig:ClassicKerrResults} (a) are presented in Fig.~\ref{Fig:ClassicKerrResults} (d). For all the t pulse power values, the estimated temporal extent of the pulses $\tau_\mathrm{p}$ results constant and equal to $\tau_\mathrm{p} = 200~$fs. The result is in good agreement with the estimated value of the temporal extent of the femtosecond pulses $\tau_\mathrm{p}$ employed in our experiment (for further information on the experimental value estimation see Note D of Suppl. Material in Ref.~\cite{Settembrini2022}).

\section{Nonlinear electric field correlation of quantum electromagnetic radiation}\label{Sect:QuantumKerrMeas}

The classical description of electro-optic and third-order nonlinear balanced detection, presented in Sec. II, assumes implicitly the coherence of the electromagnetic radiation investigated. The nonlinear balanced detection of both a coherent THz electromagnetic single mode $\vec{E}_\mathrm{THz}(t)$ and of the third order polarization $P_x^{(3)}(t+\tau)$ described in Eq.~(\ref{eq:TONPol}) is in fact based on the constant relation between the phase of the electromagnetic radiation and that of the sampling laser probe. As a consequence, measurements based on balanced ellipsometry implemented with an experimental setup as in Fig.~\ref{Fig:ClassicKerrResults} generally rely on the measurement of a large number of subsequent NIR ultrashort pulses and their averaging.

The same experimental implementation would be, however, not suited for the investigation of broadband incoherent quantum states of light, such as, for instance, thermally populated states of electromagnetic radiation and the quantum vacuum. Intuitively, the incompatibility can be attributed to the incoherent nature of the quantum electromagnetic radiation which does not exhibit a stable phase relation with the sampling laser pulse. As a result, the randomness of the measurements collected via balanced detection will lead to a null result upon direct averaging.

The same conclusion can be also obtained from a rigorous quantum mechanic description of nonlinear balanced detection, as reported in detail in Ref.~\cite{Moskalenko2015}. According to Eq.~(10) in Ref.~\cite{Moskalenko2015}, the quantum operator describing the electro-optic measurement performed by a single probing pulse $\hat{S}$ results directly proportional to the amplitude of the quantum electromagnetic field investigated $\hat{E}$. The latter, in second quantization formalism, is directly proportional to a sum of creation(destruction) operators $\hat{a}^{(\dagger)}$, $\hat{E} \propto \hat{a}+\hat{a}^{\dagger}$. In the case of thermally populated electromagnetic radiation, described as a statistical mixture state, the expectation value of the measurement operator is identically zero.

Nevertheless, the technique of electro-optic sampling has been proven to possess sufficient sensitivity to investigate the statistical properties of the electromagnetic vacuum and of its higher-order noise distribution in the MIR frequency range \cite{Riek2015,Riek2017}. The measurement of higher order noise terms, described by the operators $\hat{S}^{(2)}$, $\hat{S}^{(3)}$, etc., would in fact provide a non-vanishing result due to the presence of single mode normally and not normally ordered creation and destruction operator terms $\hat{a}^\dagger\hat{a}$, $\hat{a}\hat{a}^\dagger$.

In the THz frequency domain, the technique of electro-optic sampling combined with the use of two probing laser pulses has been proven to possess sufficient sensitivity to resolve the first-order degree coherence measurement performed on a quantum thermal state of radiation. 
The latter presents at room temperature in the THz frequency range with an average number of photons equal to only a few units within the detection bandwidth. The tunable temporal and spatial distance between the two ultrashort probes allows investigating the quantum electromagnetic state in two distinct space-time points $(\vec{r},t),~(\vec{r}+\delta \vec{r}_{\perp},t+\tau)$. The result obtained from each couple of mutually delayed pulses is then used for the direct computation of the electric field correlation function, defined in analogy with Ref.~\cite{Benea-Chelmus2019}:
\begin{equation}
    G^{(1)}(\tau, \delta \vec{r}_{\perp}) \propto \langle \{\hat{S}(t,\vec{r}),\hat{S}(t+\tau,\vec{r}+\delta \vec{r}_{\perp})\} \rangle.
    \label{eq:G1Nature}
\end{equation}
The quantum mechanics operators $\hat{S}(t,\vec{r})$, $\hat{S}(t+\tau,\vec{r}+\delta \vec{r}_{\perp})$ describe the nonlinear measurement performed by the pair of femtosecond probes at the two distinct space-time points $(\vec{r},t),~(\vec{r}+\delta \vec{r}_{\perp},t+\tau)$. The term $\{,\;\}$ indicates the anticommutator of the operators and $\langle,\;\rangle$ their expectation value.

The electric field correlation measurement technique has been shown to provide non-vanishing results even in the limit of the detectable mode's photon population tending to zero, resulting therefore as a valuable instrument for the investigation of broadband electromagnetic vacuum\cite{Benea-Chelmus2019}. The use of strongly focused Gaussian probes combined with the subcycle resolution of the employed femtosecond sampling pulses has allowed the characterization of the temporal but most importantly spatial field correlation of electromagnetic radiation in its ground state\cite{Settembrini2022}.

The incompatibility between nonlinear balanced detection and the measurement of chaotic quantum radiation can be resolved by employing a technique defined as "RF-referencing". The latter consists in referencing the measurements of each of the sampling pulses to their respective temporal adjacent pulse, which are described by the operators $\hat{S}(t+T_\mathrm{rep},\vec{r})$, $\hat{S}(t+\tau+T_\mathrm{rep},\vec{r}+\delta \vec{r}_{\perp})$. Here the quantity $T_\mathrm{rep}$ represents the time occurring between each couple of measurement pulses. As described in the Methods section of Ref.~\cite{Benea-Chelmus2019}, the electro-optic field correlation function will be given by the expression:
\begin{equation}
    G^{(1)}(\tau, \delta \vec{r}_{\perp}) = \bigl \langle \bigl( \hat{S}(t,\vec{r})-\hat{S}(t+T_\mathrm{rep},\vec{r})\bigr)\bigl(\hat{S}(t+\tau,\vec{r}+\delta \vec{r}_{\perp})-\hat{S}(t+\tau+T_\mathrm{rep},\vec{r}+\delta \vec{r}_{\perp}) \bigr)\bigr \rangle.
    \label{eq:RFReferencing}
\end{equation}
This specific measurement configuration will allow the efficient suppression of systematic coherent noise affecting each individual pulse equally, such as higher-order nonlinear balanced signal $S^{(3)}(\tau)$ described in Sec.~\ref{Chpt:Semi-ClassicalKerrExp}, and long time drifts, such as $1/f$-noise. Therefore, it confers to the nonlinear field correlation measurement technique sensitivity only to electromagnetic quantum state in which radiation presents with a coherence time smaller than $T_\mathrm{rep}$ by design.

Besides the quantum thermal radiation at THz frequencies, an additional incoherent electromagnetic light source in the experimental system is represented by the quantum vacuum at near-infrared frequencies $\hat{E}^{\mathrm{vac}.}_{x}$ polarized along the $\hat{x}$ laboratory axis, which is responsible for the generation of shot noise of each sampling laser pulse upon balanced detection~\cite{Moskalenko2015}. As the quantum vacuums characterizing the two pulses are uncorrelated, their presence does not give rise to any further contribution to the balanced correlation. However, they can influence the polarization state of the copropagating beam. 

As experimentally and theoretically demonstrated classically in Sec.~\ref{Sec:ThirdOrderClassical}, an ultrashort NIR probe with a polarization component along the $\hat{x}$ axis of the laboratory reference frame can induce a polarization change in the copropagating laser pulse, leading to the generation of a nonlinear balanced signal, described in Eq.~(\ref{eq:KerrClassicFinal}). In a similar fashion, the electromagnetic NIR vacuum $\hat{E}^{\mathrm{vac}.}_{x}$ inducing the shot noise of one sampling beam can influence the polarization of the copropagating one via four-wave mixing. The generated third-order nonlinear polarization term will induce upon balanced detection a signal correlated to the shot noise amplitude of the paired ultrashort pulse. The latter will therefore introduce a measurable nonlinear correlation term between the nonlinear balanced measurement in the two detectors.

In the following chapter, we will provide a detailed theoretical derivation of the higher-order nonlinear balanced correlation term due to the third-order nonlinear interaction of each sampling laser pulse with the quantum vacuum characterizing the copropagating probe in Sec.~\ref{Chapt:QuantumCorrKerrTheory}. The experimental characterization of the higher order nonlinear correlation term together with its experimental parameters dependence will be presented in Sec.~\ref{sec:ThirdOrderCorrExp} and Sec.~\ref{Subchapt:QuantumKerrExpPar} respectively.

\subsection{Kerr-induced nonlinear correlation of electromagnetic radiation}\label{Chapt:QuantumCorrKerrTheory}
\begin{figure}
    \centering
    \includegraphics{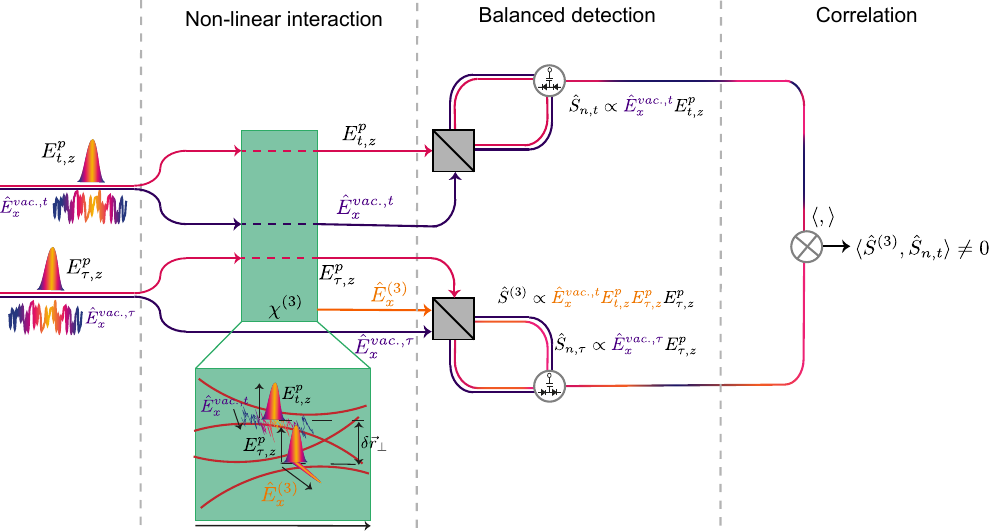}
    \caption{\textit{Quantum third order nonlinear interaction}. The four-wave mixing of the electromagnetic vacuum $\hat{E}^{\mathrm{vac}.,t}_x$ of t probing pulse with the classical fields of the femtosecond probes $\vec{E}^\mathrm{p}_{t,z},\vec{E}^\mathrm{p}_{\tau,z}$ leads to the creation of the additional time delayed probe electric field component $\hat{E}_x^{(3)}$ (in orange). Upon correlation with the quantum vacuum $\hat{E}^{\mathrm{vac}.,t}_x$ determining the shot noise of the other probe, the latter is responsible for the arising of an additional higher-order nonlinear correlation term of the form $G^{(1)}_\mathrm{Kerr}(\tau,\delta\vec{r}_{\perp})\propto \langle\hat{S}^{(3)},\hat{S}_{n,t}\rangle$, as described in Eq.~\ref{eq:QuantG1Tot}. In insert, the geometrical configuration of probing beams position and polarization inside the ZnTe crystal.}
    \label{Fig:ConceptQuantKerr}
\end{figure}

In the following, the Kerr nonlinearity caused by electromagnetic vacuum at NIR frequency able to induce a non-vanishing correlation term in the balanced ellipsometry correlation detection scheme will be proposed. It is represented schematically in Fig.~\ref{Fig:ConceptQuantKerr}. For simplicity, we will consider the influence of the electromagnetic vacuum $\hat{E}^{\mathrm{vac}.,t}_{x}$ affecting the non-time delayed t probe $E^\mathrm{p}_{t,z}(t)$ on the quantum balanced ellipsometry signal detected by the time delayed $t+\tau$ probing pulse $E^\mathrm{p}_{\tau,z}(t+\tau)$. Both probes are polarized along the $\hat{z}$ axis of the laboratory reference frame, as indicated in the pedices notation. The symmetric process obtained by exchanging the roles of time-delayed and not time-delayed probes can be shown to be formally equivalent. 

According to second quantization formalism, the amplitude of the electromagnetic vacuum at NIR frequency can be described as $\hat{E}^\mathrm{vac}=i\sum_k\sqrt{\frac{\hbar \omega_k}{2\epsilon_0n^2V}}\left[ \hat{a}_ke^{-i\omega_k t+i\vec{k}\cdot\vec{r}}- \hat{a}_k^{\dagger}e^{i\omega_k t-i\vec{k}\cdot\vec{r}} \right]$. Here the quantum operator $\hat{a}^{(\dagger)}$ describes the destruction (creation) of a NIR mode of frequency $\omega_k$ propagating with wavevector $\vec{k}$ inside the nonlinear material. In analogy with Eq.~(\ref{eq:TONPol}), the third-order nonlinear polarization arising due to the interaction of the quantum electromagnetic vacuum with the high-intensity classical electric field of the two probes can be written as:
\begin{equation}
\postdisplaypenalty=0 
\hat{P}^{(3)}_x(t+\tau)=2\epsilon_0\chi_{44}E^\mathrm{p}_{t,z}(t)E^\mathrm{p}_{\tau,z}(t+\tau)\hat{E}^{\mathrm{vac.},t}_{x}.
\label{eq:TONPolQuant}   
\end{equation}
The term $\hat{P}^{(3)}_x(t+\tau)$ reported in Eq.~(\ref{eq:TONPolQuant}) serves as a source term for the generation of the additional polarization component of the temporal delayed $t+\tau$ laser pulse $\hat{E}^{(3)}(t+\tau)$, as indicated in Fig.~\ref{Fig:ConceptQuantKerr}. The latter induces the balanced nonlinear signal at the photodetector described via the quantum mechanic operator $\hat{S}^{(3)} (t+\tau, \delta \vec{r}_{\perp})$. It can be expressed as (for detailed derivation, see Sec. II A of Suppl. Material):
\begin{equation}
\begin{aligned}
&\hat{S}^{(3)} (t+\tau, \vec{r}+\delta \vec{r}_{\perp}) = \frac{1}{2} c n\epsilon_0 \int_0^{+\infty} d\omega \frac{\eta(\omega)}{\hbar \omega}\int d^2\vec{r}_{\perp}|E^\mathrm{p}_{\tau,z}(\zeta)|^2\left[i\;\frac{\hat{E}^{(3)}}{E^\mathrm{p}_{\tau,z}(\zeta)}+\mathrm{h.c.}\right]\\    
&= -\frac{ 2\chi_{44} E^\mathrm{p}_{t} N_{\tau}\omega_c }{3cn}\int d^2\vec{r}_{\perp} g_0^2(\vec{r}-\frac{\delta \vec{r}_{\perp}}{2})g_0(\vec{r}+\frac{\delta \vec{r}_{\perp}}{2}) \sum_{\tilde{k}} \sqrt{\frac{\hbar \omega_{\tilde{k}}}{2\epsilon_0\epsilon_rV}} \left[\hat{a}_{\tilde{k}}e^{-i\omega_{\tilde{k}}t+i\vec{\tilde{k}}\cdot \delta \vec{r}_{\perp}}R(\tilde{k}_y,\omega_{\tilde{k}})-\mathrm{h.c.}\right]
\label{eq:QuantumKerrTauLine}
\end{aligned}
\end{equation}
Here the functions $g_0(\vec{r}\pm\frac{\delta\vec{r}_{\perp}}{2})$ represent the Gaussian spatial distribution of the two probing pulses centered symmetrically at $\pm\frac{\delta\vec{r}_{\perp}}{2}$ with respect to the origin of the laboratory reference frame (at the crystal center). $N_{\tau}$ indicates the number of photons of the $t+\tau$ probe effectively measured at the photodetector. The function $R(\tilde{k}_y,\omega_{\tilde{k}})$ represents the responsivity of the third-order nonlinear interaction, which is dependent on the phase matching condition of the near-infrared modes with component $\tilde{k}_y$ along the pulses propagation direction and frequency $\omega_{\tilde{k}}$. 

The quantum vacuum at near-infrared frequencies $\hat{E}^{\mathrm{vac}.,t}_{x}$ is responsible for the generation of shot noise associated with its photodetection, as also appearing directly in Ref. ~\cite{Moskalenko2015}. The measurement of the latter is described through the quantum mechanics operator $\hat{S}_{n,t}$:
\begin{align}
\begin{split}
\hat{S}_{n,t(\tau)} &= \frac{1}{2} c n\epsilon_0 \int_{0}^{+\infty} d\omega \frac{\eta(\omega)}{\hbar \omega}\int d^2\vec{r}_{\perp}\left[ (E^\mathrm{p}_{t (\tau),z}(\zeta))^* i\hat{E}_{x}^{\mathrm{vac}.,t(\tau)}+\mathrm{h.c.}\right]\\
&=-\frac{N_t\omega_c}{E^\mathrm{p}_{t,z}}\sum_{\tilde{k}^{\prime}} \sqrt{\frac{\hbar \omega_{\tilde{k}^{\prime}}}{2\epsilon_0\epsilon_r V}}\left[ \hat{a}_{\tilde{k}^{\prime}}e^{-i\omega_{\tilde{k}^{\prime}}t+i\tilde{k}^{\prime}_x O^i_{t}}-\mathrm{h.c.}\right] f(\omega_{\tilde{k}^{\prime}})\Gamma(\tilde{k}_x^{\prime},\tilde{k}_z^{\prime}).
\label{eq:QuantumNoiseT}
\end{split}
\end{align}

The functions $f(\omega_{\tilde{k}^{\prime}})$ and $\Gamma(\tilde{k}_x^{\prime},\tilde{k}_z^{\prime})$ indicate the detection responsivity for a single electromagnetic quantum vacuum mode characterized by frequency $\omega_{\tilde{k}^{\prime}}$ and transverse wavevectors components $\tilde{k}_x^{\prime},\tilde{k}_z^{\prime}$ respectively. An equivalent expression can be derived for the shot noise operator $\hat{S}_{n,\tau}$ of the delayed $t+\tau$ probe. 

Following the derivations present in literature, the quantum mechanics operator $\hat{S}_\mathrm{tot.}$ describing the complete result of the balanced ellipsometry measurement performed by the two femtosecond pulses in the space-time points $(t,\vec{r}),(t+\tau,\vec{r}+\delta \vec{r}_{\perp})$ individually can be written as a sum of the following contributions:
\begin{subequations}
\begin{equation}
\postdisplaypenalty=0 
\hat{S}_\mathrm{tot.}(t,\vec{r}) = \hat{S}_\mathrm{eo}(t,\vec{r}) + \hat{S}^{(3)}(t,\vec{r}) +\hat{S}_{n,t},
\label{eq:TOTQuantT}   
\end{equation}
\begin{equation}
\hat{S}_\mathrm{tot.}(t+\tau,\vec{r}+\delta \vec{r}_{\perp}) = \hat{S}_\mathrm{eo}(t+\tau,\vec{r}+\delta \vec{r}_{\perp})+\hat{S}^{(3)}(t+\tau,\vec{r}+\delta \vec{r}_{\perp}) +\hat{S}_{n,\tau}.
\label{eq:TOTQuantTau}   
\end{equation}
\end{subequations}
Here the operators $\hat{S}_\mathrm{eo}(t,\vec{r}),\hat{S}_\mathrm{eo}(t+\tau,\vec{r}+\delta \vec{r}_{\perp})$ are related to the electro-optic measurement of THz electromagnetic radiation, as defined in Ref.~\cite{Benea-Chelmus2019}. Rewriting Eq.~(\ref{eq:RFReferencing}) as a function of the quantum operators defined in Eq.~(\ref{eq:TOTQuantT}) and Eq.~(\ref{eq:TOTQuantTau}), the final experimental result will be given by the expectation value of the quantum operator on a thermally populated radiation state. Taking into consideration only the lower nonlinear contribution terms, the result will read:
\begin{equation}
    \begin{aligned}
    &G^{(1)}(\tau,\delta\vec{r}_{\perp}) = \bigl \langle \bigl\{ \hat{S}_\mathrm{tot.}(t,\vec{r}),\hat{S}_\mathrm{tot.}(t+\tau,\vec{r}+\delta \vec{r}_{\perp}\bigr\}\bigr \rangle\\
    &= \bigl \langle \bigl\{ \hat{S}_\mathrm{eo}(t,\vec{r}),\hat{S}_\mathrm{eo}(t+\tau,\vec{r}+\delta \vec{r}_{\perp})\bigr\}\bigr \rangle + \bigl \langle \bigl\{  \hat{S}^{(3)}(t,\vec{r}),\hat{S}_{n,\tau}\bigr\}\bigr \rangle + \bigl \langle \bigl\{  \hat{S}^{(3)}(t+\tau,\vec{r}+\delta\vec{r}_{\perp}),\hat{S}_{n,t}\bigr\}\bigr \rangle.
    \label{eq:QuantG1Tot}
    \end{aligned}
\end{equation}
Here, $ \bigl\{,\bigl\}$ indicates the anti-commutator of the two operators and $\bigl \langle,\bigl \rangle$ the quantum mechanics expectation value on a quantum thermal state of radiation, which can be formally described as a statistical mixture state.
According to quantum mechanics, the only non-vanishing terms upon computation of the expectation value of Eq.~(\ref{eq:QuantG1Tot}) will be terms presenting both the normally ordered $\hat{a}^{\dagger}\hat{a}$ and not normally ordered $\hat{a}\hat{a}^{\dagger}$ creation and destruction operators relative to the same electromagnetic mode. The first surviving term $\bigl \langle \bigl\{ \hat{S}_\mathrm{eo}(t,\vec{r}),\hat{S}_\mathrm{eo}(t+\tau,\vec{r}+\delta \vec{r}_{\perp})\bigr\}\bigr \rangle$ represents the electro-optic electric field correlation $G^{(1)}_\mathrm{eo.}(\tau,\delta\vec{r}_{\perp})$ of THz electromagnetic radiation, as defined in Ref.~\cite{Benea-Chelmus2019,Settembrini2022}. The additional terms $\bigl \langle \bigl\{ \hat{S}^{(3)}(t,\vec{r}),\hat{S}_{n,\tau}\bigr\}\bigr \rangle$ and $\bigl \langle \bigl\{\hat{S}^{(3)}(t+\tau,\vec{r}+\delta\vec{r}_{\perp}),\hat{S}_{n,t}\bigr\}\bigr \rangle$ are induced from the correlation of the electromagnetic vacuum responsible for the shot noise characterizing one ultrashort probe with the third order nonlinear balanced signal induced by the same quantum vacuum on the copropagating pulse. Due to wave-vector conservation, the nonlinear balanced signal detected individually by each femtosecond pulse could not arise from a four-wave mixing process induced via its own quantum vacuum. It is also important to note that the same argument cannot be applied to the quantum vacuum fluctuations at THz frequencies, due to their very long wavelength that relaxes the phase-matching condition.
  
The shot noise in the two photodetectors also results uncorrelated  $\langle\{\hat{S}_{n,t},\hat{S}_{n,\tau}\}\rangle = 0$. This result derives directly from the lack of correlation between the electromagnetic quantum vacuum characterizing the sampling probes $\hat{E}^{\mathrm{vac}.,t}_x$ and $\hat{E}^{\mathrm{vac}.,\tau}_x$. Even if stemming from the same source, the presence of a beam splitter in the optical path as illustrated in Fig.~\ref{Fig:QuantKerrResults} (a) is responsible for different destruction and creation operators of the quantum electromagnetic vacuum affecting the two probes~\cite{Guedes2022}, lifting their degeneracy.

To summarize, a simple quantum mechanical description of both the noise characterizing the two probes and of the incoherent nonlinear balanced ellipsometry measurement predicts, in addition to the term arising from the THz quantum thermal radiation, the existence of an additional correlation term $G^{(1)}_\mathrm{Kerr}(\tau,\delta\vec{r}_{\perp}) =\bigl \langle \bigl\{ \hat{S}^{(3)}(t,\vec{r}),\hat{S}_{n,\tau}\bigr\}\bigr \rangle +\bigl \langle \bigl\{\hat{S}^{(3)}(t+\tau,\vec{r}+\delta\vec{r}_{\perp}),\hat{S}_{n,t}\bigr\}\bigr \rangle$ arising from a third-order mixing between the quantum vacuum at the NIR frequency and the propagating pulse of the other line. In other words, vacuum fluctuations responsible for the shot noise on one line mix with the pulse on the other line and create a correlated noise affecting the latter after the ellipsometry measurement.

\subsection{Experimental nonlinear field correlation}\label{sec:ThirdOrderCorrExp}

\begin{figure*}
    \centering
    \includegraphics{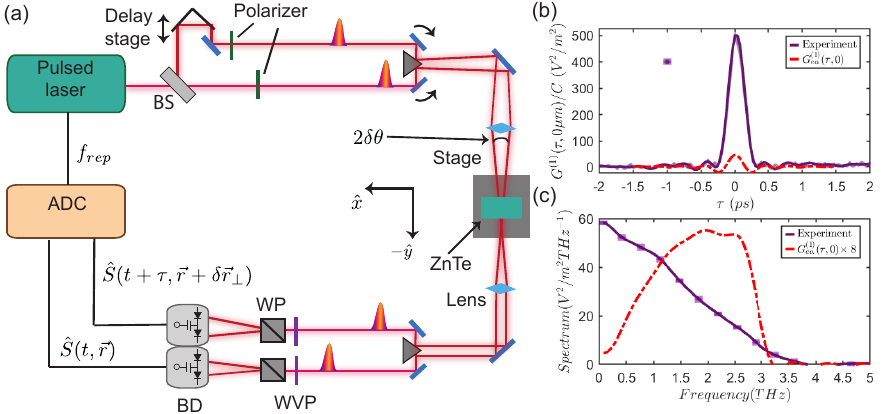}
    \caption{\textit{Experimental quantum higher order nonlinear correlation}. (a) The experimental setup for nonlinear electric field correlation measurements allows the control of both the temporal and spatial distance between probes. The latter is controlled via a couple of piezo mirrors, which confer a relative propagation angle of $2\delta\theta$ to the two probes. The acquisition of the experimental measurement points is performed at the repetition rate of the femtosecond laser $f_\mathrm{rep} = 80~$MHz. BD = Balanced detectors, WP = Wollaston prism, WVP = Quarter waveplate. (b),(c) Experimental electric field correlation measurement results performed with a beams' spatial separation of $\delta\vec{r} = 0~\upmu$m in time (b) and frequency domain (c) (in faded purple). For better visualization, the data have been filtered with a Kaiser windowing function (in solid purple). For comparison, the numerically simulated result for the electro-optic field correlation $G_\mathrm{eo.}^{(1)}(\tau,\delta \vec{r}_{\perp} = 0)$ of THz thermal radiation at 300 K is reported in dashed-dotted lines. In both figures, the experimental uncertainty indicates the $2\sigma$ confidence interval.}
    \label{Fig:QuantKerrResults}
\end{figure*}

The investigation of quantum field correlation measurement on the thermally populated electromagnetic state has been carried out using the experimental setup reported in Fig.~\ref{Fig:QuantKerrResults} (a). The system is similar to the one presented in Fig.~\ref{Fig:ClassicKerrResults} (a), but in addition to the temporal distance between the two probing ultrashort pulses, also their spatial distance $\delta \vec{r}_{\perp}$ in the transverse plane can be controlled via a symmetric couple of piezo mirrors. The change in the propagation angle $\delta \theta$ with respect to the femtosecond pulses propagation direction $\hat{y}$ will be translated by the lens system into a spatial separation $\delta \vec{r}_{\perp}=f \delta \theta$. 
The two femtosecond probing pulses are focused by the lens onto the detection crystal, where they will sample the same electromagnetic state at two different space-time points. Afterward, they will be individually analyzed via balanced ellipsometry, where the acquisition is performed at the repetition rate of the laser $f_\mathrm{rep.}$. The recording of the results obtained from each couple of near-infrared pulses allows the real-time computation of the electric field correlation function as described in Eq.~(\ref{eq:RFReferencing}).

The experimental nonlinear field correlation measurement obtained using the setup presented in Fig.~\ref{Fig:QuantKerrResults} (a) with overlapping sampling beams of Gaussian beam waist $w_0 = 10~\upmu$m, temporal extent $\tau_\mathrm{p} = 200~$fs and peak electric field amplitude of $E^\mathrm{p}_z = 10~$MV$/$m are reported in Fig.~\ref{Fig:QuantKerrResults} (b) and (c) in time and frequency domain respectively. As shown in figure, the results measured using the specified set of experimental parameters for perfectly overlapping sampling beams $\delta\vec{r}_{\perp} = 0$ differ significantly both in amplitude and most importantly in spectral content from the expected electro-optic THz field correlation $G_\mathrm{eo.}^{(1)}(\tau,\delta \vec{r}_{\perp} = 0)$ generated via blackbody emission from the environment at $300~$K, also shown for comparison. The latter has been numerically estimated using the expression reported in Eq. (2) of Ref.~\cite{Benea-Chelmus2019} and the experimental parameters of the femtosecond probing pulses employed in the experiment. 

In order to compare the measured phase correlation between the sampling probes obtained via balanced ellipsometry with the one due to electro-optic field correlation $G_\mathrm{eo.}^{(1)}(\tau,\delta\vec{r}_{\perp})$ induced by THz thermal radiation, all experimental measurements have been normalized by the constant $C = \frac{r_{41}n^3l\omega_\mathrm{p} I_{\mathrm{p,}t}I_{\mathrm{p,}\tau}}{c}$, provided in literature \cite{Moskalenko2015,Benea-Chelmus2019,Settembrini2022}. 
Here, $r_{41}$ represents the electro-optic coefficient of the nonlinear ZnTe crystal and $I_{\mathrm{p,}t}$, $I_{\mathrm{p,}\tau}$ the femtosecond probes intensity impinging on the balanced photodetectors. The experimental field correlations have been moreover filtered with a Kaiser windowing function, in order to reduce the presence of noise-induced artefacts (for more informations see Supplementary Note 3 of Ref.~\cite{Settembrini2022}).
 
As reported in Fig.~\ref{Fig:QuantKerrResults} (b), the amplitude of the experimental signal presents a peak-to-peak amplitude of around $G^{(1)}_\mathrm{pp}(\tau, 0~\upmu$m$) = 500~$V$^2/$m$^2$, which results approximately ten times larger than the expected correlation of THz photons $G^{(1)}_\mathrm{eo,pp.}(\tau, 0~\upmu$m$) = 55~$V$^2/$m$^2$. The two curves present significant differences as well in frequency content, as it is shown by their Fourier transformation in Fig.~\ref{Fig:QuantKerrResults} (c). While the spectral content of the numerically simulated electric field correlation 
 $G^{(1)}_\mathrm{eo.}(\tau, 0~\upmu$m$)$ depends on the phase matching properties of THz electromagnetic modes in ZnTe and therefore presents mostly contributions from higher frequency components around $2~$THz, the experimental spectrum derived from Fig.~\ref{Fig:QuantKerrResults} (b) presents maximum contributions in the low-frequency components regions with a maximum centered at zero frequency.

\subsection{Experimental nonlinear field correlation dependence on experimental parameters}\label{Subchapt:QuantumKerrExpPar}

The nature of the measured nonlinear correlation has been investigated by studying its dependence on experimental parameters such as temperature, crystal length, and probing pulses' wavelength, power, and spatial distance $\delta \vec{r}_{\perp}$. In order to compare the nonlinear correlation measurement to the numerically predicted results induced by electro-optic detection of thermal THz radiation $G^{(1)}_\mathrm{eo.}(\tau,\delta \vec{r}_{\perp})$, the experimental measurements presented in this section have also been normalized by the electro-optic correlation constant $C$  and filtered as defined in Sec.~\ref{Sect:QuantumKerrMeas}.

\begin{figure}
    \centering
    \includegraphics{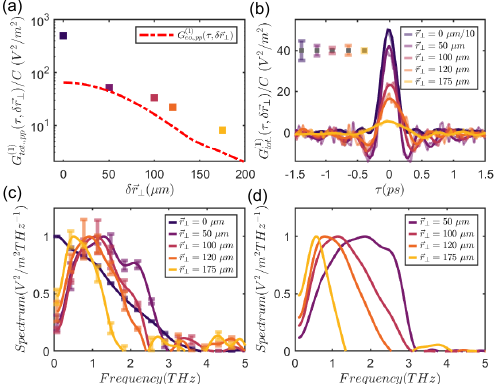}
    \caption{\textit{Quantum higher order nonlinear correlations as a function of beam separation}. (a) Dependence of the peak-to-peak amplitude of higher order nonlinear correlation term $G^{(1)}_\mathrm{pp.}(\tau,\delta\vec{r}_{\perp})$ (in log scale) as a function of the transverse beam separation $\delta\vec{r}_{\perp}$. For comparison, also the peak-to-peak amplitude of the electro-optic field correlation of THz modes $G^{(1)}_\mathrm{eo.,pp}(\tau,\delta\vec{r}_{\perp})$ is reported (dash-dotted line). (b,c) Experimental nonlinear correlation measurements in time (b) and frequency domain (c) for increasing $\delta\vec{r}_{\perp}$. All errorbars represent the $2\sigma$ confidence interval. (d) Numerically computed  electro-optic electric field correlation $G^{(1)}_\mathrm{eo.}(\tau,\delta\vec{r}_{\perp})$ at T = $300~$K for spatial beams separation $\delta\vec{r}_{\perp}$ reported in (b) and (c) for non-overlapping beams.}
    \label{Fig:QuantKerrSpatial}
\end{figure}

As derived theoretically in Sec. III A and more in detail in Sec. II A of the Suppl. Material, the higher-order nonlinear correlation term $G^{(1)}_\mathrm{Kerr}(\tau,\delta\vec{r}_{\perp}) =\bigl \langle \bigl\{\hat{S}^{(3)}(t+\tau,\vec{r}+\delta\vec{r}_{\perp}),\hat{S}_{n,t}\bigr\}\bigr \rangle$ and the THz blackbody induced field correlation $G^{(1)}_\mathrm{eo.}(\tau,\delta\vec{r}_{\perp})$ will present strongly different temporal coherence characteristics, which will strongly depend on the sampling beams transverse separation $\delta\vec{r}_{\perp}$. This dependence has been investigated experimentally. The 
measured nonlinear correlation $G^{(1)}_\mathrm{tot.,pp}(\tau,\delta\vec{r}_{\perp})$  amplitude dependence on the relative distance between the sampling probes $\delta \vec{r}_{\perp}$ is reported both in time and frequency domain in Fig.~\ref{Fig:QuantKerrSpatial}.

In Fig.~\ref{Fig:QuantKerrSpatial} (a) the total peak-to-peak amplitude of the measured nonlinear correlation $G^{(1)}_\mathrm{tot.,pp.}(\tau,\delta\vec{r}_{\perp})$ is reported as a function of the relative transverse distance between the sampling probes $\delta\vec{r}_{\perp}$. As the latter increases, the nonlinear correlation significantly decreases in amplitude from a value of $G^{(1)}_\mathrm{tot.,pp.}(\tau,\delta\vec{r}_{\perp}) = 500~$V$^2/$m$^2$ in the case of perfectly overlapping sampling beams $\delta\vec{r}_{\perp} = 0~\upmu \mathrm{m}$ to a value of $G^{(1)}_\mathrm{tot.,pp.}(\tau,\delta\vec{r}_{\perp}) = 8.2~$V$^2/$m$^2$ for a transverse beam separation of $\delta\vec{r}_{\perp} = 175~\upmu \mathrm{m}$. As shown in figure by comparison with the numerically simulated result, for spatially overlapping sampling laser probes the measured correlation is dominated by higher-order nonlinear term $G^{(1)}_\mathrm{Kerr}(\tau,\delta\vec{r}_\perp)$, while for sampling beams with a spatial transverse distance equal several times their diameters 
the experimental results are well predicted by the electro-optic field correlation of THz thermal radiation $G^{(1)}_\mathrm{eo.}(\tau,\delta\vec{r}_{\perp})$. A similar behavior can be observed also employing different nonlinear ZnTe detection crystals of various lengths, as shown in Sec. III in the supplementary material.

As expected, the experimental correlation measurements performed for different probing beams' distances present significant differences not only in amplitude but in the extent of their temporal coherence as well, as it is reported in  Fig.~\ref{Fig:QuantKerrSpatial} (b). The experimental result measured in the case of $\delta\vec{r}_{\perp} = 0~\upmu \mathrm{m}$ preserves its temporal coherence for a time of $0.5~$ps. An increase in spatial distance between sampling probes $\delta \vec{r}_{\perp}$ correlates on the other hand with an increased temporal extent of the nonlinear correlation, which is preserved for over $2~$ps for all the measurements. This difference is reflected in the spectral content of the sampled broadband radiation in the case of overlapping and non-overlapping sampling pulses, which are reported in Fig.~\ref{Fig:QuantKerrSpatial} (c). As it can be seen in figure, the spectral content of the nonlinear correlation measured with $\delta \vec{r}_{\perp} = 0~\upmu \mathrm{m}$ presents a large bandwidth of $3.5~$THz with the majority of the contributions given by lower frequency modes around $0~$THz, in good agreement with a major contribution stemming from the higher-order nonlinear correlation $G^{(1)}_\mathrm{Kerr}(\tau,\delta\vec{r}_{\perp})$. On the other hand, the experimental result obtained with the sampling beams non-overlapping inside the nonlinear material $\delta \vec{r}_{\perp} > 50~\upmu \mathrm{m}$ show low spectral contributions from the low-frequency region. Moreover, their spectral contents present a varying bandwidth and peak contribution frequency, which decrease from a value of $3~$THz and $1.5~$THz respectively for the measurement performed with $\delta \vec{r}_{\perp} = 50~\upmu \mathrm{m}$ to values of $1.75~$THz and $0.75~$THz for the nonlinear correlation measured with $\delta \vec{r}_{\perp} = 175~\upmu \mathrm{m}$.
The decrease in bandwidth as well as the redshift of the peak contribution frequency for nonlinear correlation measured with non-spatially overlapping probes ($\delta \vec{r}_{\perp} > 50~\upmu \mathrm{m}$) presents a good agreement with the numerical results for electro-optic field correlation of thermal THz radiation, reported in Fig.~\ref{Fig:QuantKerrSpatial} (d). As it is shown in figure, the numerical THz electro-optic correlation estimated for different transverse beams spatial separations $\delta \vec{r}_{\perp}$ present almost identical bandwidths to the experimental results and similar redshift in the peak contribution frequency, which decrease from a value of $3.2~$THz and $2~$THz respectively for the measurement performed with $\delta \vec{r}_{\perp} = 50~\upmu \mathrm{m}$ to values of $1.4~$THz and $0.6~$THz for the nonlinear correlation measured with $\delta \vec{r}_{\perp} = 175~\upmu \mathrm{m}$. The redshift of the peak contribution frequency of the THz electro-optic correlation can be intuitively explained in the spatial domain with electromagnetic modes of wavelength equal or smaller than twice the spatial transverse distance between the sampling pulses providing negative or null contributions to the measured field correlation.

\begin{figure*}[h!]
    \centering
    \includegraphics{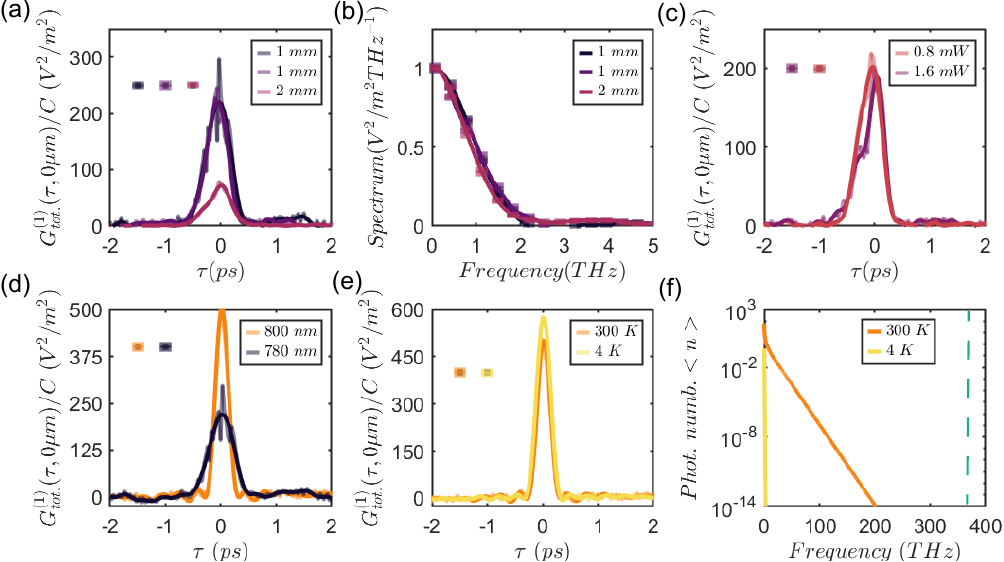}
    \caption{\textit{Experimental parameters dependence}. (a,b) Higher order nonlinear correlation measurement in time (a) and frequency domain (b) employing two different 1mm and a 2 mm long $\langle 110 \rangle$-cut ZnTe detection crystal. The experimental parameters used are: $P_t=P_{\tau}=0.8~$mW, $\tau_\mathrm{p} = 200~$fs, $\lambda = 780~$nm. (c) Third-order nonlinear correlation measurement for different sampling laser pulses' power. Both measurements have been performed using a 1 mm long ZnTe crystal, $\tau_\mathrm{p} = 200~$fs and $\lambda = 780~$nm. (d). Experimental nonlinear correlation measured employing radiation of $\lambda = 800~$nm, $780~$nm. The measurements have been performed using the same $\langle 110 \rangle$-cut ZnTe crystal. (e) Effect of the temperature of the blackbody radiation detected on the higher order nonlinear correlation $G^{(1)}_\mathrm{tot.}(\tau, 0~\upmu\mathrm{m})$. (f) Average number of photons per mode $\langle n \rangle$ according to Planck's radiation law at different temperatures of $300~$K (orange) and $4~$K (yellow). The central frequency of the ultrashort probing pulse employed in the experiment $\omega_c = 375~$THz is reported in blue dashed line. The errorbars in all the experimental measurements in figure represent the $2\sigma$ confidence interval.}
    \label{Fig:QuantKerrParameters}
\end{figure*}

In order to corroborate the hypothesis on the origin of the higher-order nonlinear correlation from a vacuum-assisted four-wave mixing of the overlapping probing pulses, a further analysis of its dependence on the experimental parameters has been performed. The experimental results obtained as a function of different crystals, environment temperature and probes' power and wavelength are reported in Fig.~\ref{Fig:QuantKerrParameters}. All the measurements shown have been performed with perfectly overlapping sampling beams $\delta\vec{r}_{\perp} = 0~\upmu\mathrm{m}$.

The higher-order nonlinear correlation measurements obtained employing different lengths of the $\langle 110 \rangle$-cut ZnTe detection crystal are reported in Fig.~\ref{Fig:QuantKerrParameters} (a). As it can be seen, the normalized nonlinear correlation measurement performed using two distinct 1 mm long and a 2 mm long ZnTe detection crystals are all characterized by the presence of a higher order correlation term, whose amplitude decreases from a peak value of $G^{(1)}_\mathrm{tot.}(\tau, 0~\upmu\mathrm{m}) = 230~$V$^2/$m$^2$ for the 1 mm long crystal to a value of $G^{(1)}_\mathrm{tot.}(\tau, 0~\upmu\mathrm{m}) = 75~$V$^2/$m$^2$ for the 2 mm long crystal. All the measurements present however similar temporal coherence, which is preserved for a period of $0.75~$ps. Their common origin in both crystals is further highlighted by their comparison in frequency domain, presented in Fig.~\ref{Fig:QuantKerrParameters} (b). All experimental measurements present exactly the same spectral content, with a bandwidth of around $2~$THz and the majority of the contributions originating from the lower frequency region around $0~$THz. The latter is in good agreement with a nonlinear correlation signal generated via third-order nonlinear mixing of the probes mediated via the electromagnetic vacuum, as described in Eq.~(33) of Supp. Material in Sec. II A. In contrast, a signal generated from the electro-optic correlation of the THz thermal background field would have significantly depended on the phase-matching condition and therefore on the length of the crystal. 

A further indication of the third-order nonlinear origin of the measured signal is shown in Fig.~\ref{Fig:QuantKerrParameters} (c), where the measurements performed employing a sampling pulses probing powers of $P_t=P_{\tau}=1.6~$mW and $P_t=P_{\tau}=0.8~$mW respectively are shown. Given the power normalization introduced by the use of the constant C, the experimental higher-order nonlinear correlation measurements appear to be directly proportional to the product of the power of the sampling probes $P_t$, $P_{\tau}$. This result is once more in good agreement with the expected power dependency of a third-order induced nonlinear correlation, as shown in Eq.~(33) of Suppl. Material in Sec.II A.

The dependence of the balanced ellipsometry correlation amplitude on the wavelength of the sampling radiation $\lambda$ is reported in Fig.~\ref{Fig:QuantKerrParameters} (d). As it can be seen in figure, the use of a longer wavelength of near-infrared radiation $\lambda = 800~$nm leads to a higher order nonlinear correlation of peak-to-peak amplitude of around $G^{(1)}_\mathrm{tot.}(\tau, 0~\upmu\mathrm{m}) = 500~$V$^2/$m$^2$, twice as large with respect to the result obtained using red-shifted sampling probes centered at $\lambda = 780~$nm.

The influence of the temperature $T$ of the thermal radiation to which the ZnTe detection crystal is exposed and, as a consequence, of the average number of photons per mode $\langle n \rangle$ populating the incoherent quantum radiation responsible for the nonlinear correlation is reported in Fig.~\ref{Fig:QuantKerrParameters} (e). As shown in figure, the change in temperature $T$ of the blackbody radiation measured does not affect significantly the amplitude of the higher order nonlinear correlation amplitude, which increases from a peak-to-peak value of $G^{(1)}_\mathrm{tot.}(\tau, 0~\upmu\mathrm{m}) = 500~$V$^2/$m$^2$ at $T = 300~$K to a value of $G^{(1)}_\mathrm{tot.}(\tau, 0~\upmu\mathrm{m}) = 580~$V$^2/$m$^2$ at $T = 4~$K. The experimental result is in good agreement with a higher-order correlation term generated via the electromagnetic vacuum at NIR frequencies. As it can be seen in Fig.~\ref{Fig:QuantKerrParameters} (f), the average number of photons per mode $\langle n \rangle$ at NIR frequencies, given by Planck's radiation law, in the frequency range around the central frequency of the infrared probing pulse $\omega_c = 375~$THz remains below a value of $\langle n \rangle < 10^{-14}$ for both temperatures.
\section{Conclusion}
 In this work, we have shown how third-order nonlinearity at NIR frequencies can affect the balanced ellipsometry measurement performed with ZnTe, both on the signal measured by the individual pulses and most importantly on their correlation.
 
 In Sec.~\ref{Chapt:Semi-ClassKerrTheory} we have derived theoretically the properties in time and in frequency domain of the third-order coherent signal induced on each of the sampling pulses due to their mutual interaction in the nonlinear material. The predictions have been corroborated by the experimental results reported in Sec.~\ref{Chpt:Semi-ClassicalKerrExp}. 
 
 In a balanced field correlation measurement, third-order nonlinearities have also been shown to play a significant role in the generation of correlation between the incoherent signals measured by the two sampling pulses. As proven theoretically in Sec.~\ref{Chapt:QuantumCorrKerrTheory}, the electromagnetic vacuum at NIR frequencies characterizing the noise of one pulse can interact with the copropagating one inside the nonlinear medium, leading to the change of its polarization. Upon measurement, the vacuum-induced change in polarization of one femtosecond pulse will therefore result correlated with the noise of the copropagating pulse, which is induced by the same quantum vacuum. 
 
 The experimental characterization of the vacuum-assisted third-order nonlinear correlation has been reported in Sec.~\ref{Subchapt:QuantumKerrExpPar}. Here, the analysis of the spatial nonlinear correlation of incoherent radiation clearly indicates the presence of two different regimes. Whilst for spatially overlapping sampling beams the nonlinear correlation result is dominated by the four-wave mixing due to the quantum vacuum at NIR frequencies, for increasing distance between the sampling pulses the experimental results return to a good agreement with electro-optic field correlation of THz thermal radiation. 
 A detailed analysis of the experimental parameters dependence of the third-order nonlinear correlation further verifies its connection to the electromagnetic vacuum at NIR frequencies, given its independence on temperature, power dependence and its presence with similar features in different crystals. All our experimental findings present a good agreement with the predicted measurement parameters dependence of the responsivity of the higher-order nonlinear correlation, derived in Sec. II A and B of the supplementary material. The qualitative analysis of the latter, moreover, clearly implies the absence of significant contributions from vacuum-induced third-order nonlinear correlation due to quantum vacuum at NIR frequencies in the electro-optic field correlation measurements of THz thermal radiation presented in our previous work \cite{Benea-Chelmus2019, Settembrini2022}. 
 
 Because two-beam correlation experiments, in contrast to single-beam studies~\cite{Riek2015,Riek2017}, allow the identification of the frequency components contributing to the correlation signal, they also have allowed to isolate the contribution from the higher order non-linear signal that is inevitably mixed into the electro-optic sampling of the THz radiation. Furthermore, it shows that measuring two nearby spatial locations with beams that do not physically overlap is a very efficient strategy to remove such contribution while increasing the detection sensitivity and allowing further studies of the quantum vacuum~\cite{Settembrini2022}.     

\begin{acknowledgments}
The experimental work was funded by the Swiss National Science Foundation (grant $200020~ 192330 /1$) and the National Centre of Competence in Research Quantum Science and Technology (NCCR QSIT) (grant 51NF40-185902) (F.F.S, A.H.). We acknowledge the mechanical workshop at ETHZ. We acknowledge the contributions of Dr. Cristina Ileana Benea-Chelmus for their contributions to previous work, Dr. E. Mavrona for the extraction of the refractive index of ZnTe. We thank Prof. Denis Seletskiy for fruitful discussion.
\end{acknowledgments}

\section*{Authors declaration}

\subsection*{Conflict of interest}

The authors declare no conflict of interest.

\subsection*{Authors contribution}

J.F. and F.F.S. conceived the idea for the experiment and its theoretical interpretation. F.F.S and A.H. conducted the measurements. The data analysis was primarily performed by F.F.S and their results were interpreted by F.F.S, A.H and J.F. The theoretical framework was developed by F.F.S. J.F. was the scientific supervisor of this work. The manuscript was written through contributions from all authors. All authors have given approval to the final version of the manuscript.

\section*{Data Availability Statement}

The experimental data supporting the findings of this study are available from the corresponding author upon reasonable request.

\bibliography{Bibliography}

\providecommand{\noopsort}[1]{}\providecommand{\singleletter}[1]{#1}%
\begin{thebibliography}{29}%
\makeatletter
\providecommand \@ifxundefined [1]{%
 \@ifx{#1\undefined}
}%
\providecommand \@ifnum [1]{%
 \ifnum #1\expandafter \@firstoftwo
 \else \expandafter \@secondoftwo
 \fi
}%
\providecommand \@ifx [1]{%
 \ifx #1\expandafter \@firstoftwo
 \else \expandafter \@secondoftwo
 \fi
}%
\providecommand \natexlab [1]{#1}%
\providecommand \enquote  [1]{``#1''}%
\providecommand \bibnamefont  [1]{#1}%
\providecommand \bibfnamefont [1]{#1}%
\providecommand \citenamefont [1]{#1}%
\providecommand \href@noop [0]{\@secondoftwo}%
\providecommand \href [0]{\begingroup \@sanitize@url \@href}%
\providecommand \@href[1]{\@@startlink{#1}\@@href}%
\providecommand \@@href[1]{\endgroup#1\@@endlink}%
\providecommand \@sanitize@url [0]{\catcode `\\12\catcode `\$12\catcode `\&12\catcode `\#12\catcode `\^12\catcode `\_12\catcode `\%12\relax}%
\providecommand \@@startlink[1]{}%
\providecommand \@@endlink[0]{}%
\providecommand \url  [0]{\begingroup\@sanitize@url \@url }%
\providecommand \@url [1]{\endgroup\@href {#1}{\urlprefix }}%
\providecommand \urlprefix  [0]{URL }%
\providecommand \Eprint [0]{\href }%
\providecommand \doibase [0]{https://doi.org/}%
\providecommand \selectlanguage [0]{\@gobble}%
\providecommand \bibinfo  [0]{\@secondoftwo}%
\providecommand \bibfield  [0]{\@secondoftwo}%
\providecommand \translation [1]{[#1]}%
\providecommand \BibitemOpen [0]{}%
\providecommand \bibitemStop [0]{}%
\providecommand \bibitemNoStop [0]{.\EOS\space}%
\providecommand \EOS [0]{\spacefactor3000\relax}%
\providecommand \BibitemShut  [1]{\csname bibitem#1\endcsname}%
\let\auto@bib@innerbib\@empty
\bibitem [{\citenamefont {Chang}\ \emph {et~al.}(2014)\citenamefont {Chang}, \citenamefont {Vuleti{\'{c}}},\ and\ \citenamefont {Lukin}}]{Chang2014}%
  \BibitemOpen
  \bibfield  {author} {\bibinfo {author} {\bibfnamefont {D.~E.}\ \bibnamefont {Chang}}, \bibinfo {author} {\bibfnamefont {V.}~\bibnamefont {Vuleti{\'{c}}}},\ and\ \bibinfo {author} {\bibfnamefont {M.~D.}\ \bibnamefont {Lukin}},\ }\bibfield  {title} {\bibinfo {title} {Quantum nonlinear optics --- photon by photon},\ }\href {https://doi.org/10.1038/nphoton.2014.192} {\bibfield  {journal} {\bibinfo  {journal} {Nature Photonics}\ }\textbf {\bibinfo {volume} {8}},\ \bibinfo {pages} {685} (\bibinfo {year} {2014})}\BibitemShut {NoStop}%
\bibitem [{\citenamefont {Kimble}(2008)}]{Kimble2008}%
  \BibitemOpen
  \bibfield  {author} {\bibinfo {author} {\bibfnamefont {H.~J.}\ \bibnamefont {Kimble}},\ }\bibfield  {title} {\bibinfo {title} {The quantum internet},\ }\href {https://doi.org/10.1038/nature07127} {\bibfield  {journal} {\bibinfo  {journal} {Nature}\ }\textbf {\bibinfo {volume} {453}},\ \bibinfo {pages} {1023} (\bibinfo {year} {2008})}\BibitemShut {NoStop}%
\bibitem [{\citenamefont {Knill}\ \emph {et~al.}(2001)\citenamefont {Knill}, \citenamefont {Laflamme},\ and\ \citenamefont {Milburn}}]{Knill2001}%
  \BibitemOpen
  \bibfield  {author} {\bibinfo {author} {\bibfnamefont {E.}~\bibnamefont {Knill}}, \bibinfo {author} {\bibfnamefont {R.}~\bibnamefont {Laflamme}},\ and\ \bibinfo {author} {\bibfnamefont {G.~J.}\ \bibnamefont {Milburn}},\ }\bibfield  {title} {\bibinfo {title} {A scheme for efficient quantum computation with linear optics},\ }\href {https://doi.org/10.1038/35051009} {\bibfield  {journal} {\bibinfo  {journal} {Nature}\ }\textbf {\bibinfo {volume} {409}},\ \bibinfo {pages} {46} (\bibinfo {year} {2001})}\BibitemShut {NoStop}%
\bibitem [{\citenamefont {Schlawin}(2017)}]{Schlawin2017}%
  \BibitemOpen
  \bibfield  {author} {\bibinfo {author} {\bibfnamefont {F.}~\bibnamefont {Schlawin}},\ }\bibfield  {title} {\bibinfo {title} {Entangled photon spectroscopy},\ }\href {https://doi.org/10.1088/1361-6455/aa8a7a} {\bibfield  {journal} {\bibinfo  {journal} {Journal of Physics B: Atomic, Molecular and Optical Physics}\ }\textbf {\bibinfo {volume} {50}},\ \bibinfo {pages} {203001} (\bibinfo {year} {2017})}\BibitemShut {NoStop}%
\bibitem [{\citenamefont {Giovannetti}\ \emph {et~al.}(2011)\citenamefont {Giovannetti}, \citenamefont {Lloyd},\ and\ \citenamefont {Maccone}}]{Giovannetti2011}%
  \BibitemOpen
  \bibfield  {author} {\bibinfo {author} {\bibfnamefont {V.}~\bibnamefont {Giovannetti}}, \bibinfo {author} {\bibfnamefont {S.}~\bibnamefont {Lloyd}},\ and\ \bibinfo {author} {\bibfnamefont {L.}~\bibnamefont {Maccone}},\ }\bibfield  {title} {\bibinfo {title} {Advances in quantum metrology},\ }\href {https://doi.org/10.1038/nphoton.2011.35} {\bibfield  {journal} {\bibinfo  {journal} {Nature Photonics}\ }\textbf {\bibinfo {volume} {5}},\ \bibinfo {pages} {222} (\bibinfo {year} {2011})}\BibitemShut {NoStop}%
\bibitem [{\citenamefont {Miller}(2010)}]{Miller2010}%
  \BibitemOpen
  \bibfield  {author} {\bibinfo {author} {\bibfnamefont {D.~A.~B.}\ \bibnamefont {Miller}},\ }\bibfield  {title} {\bibinfo {title} {Are optical transistors the logical next step?},\ }\href {https://doi.org/10.1038/nphoton.2009.240} {\bibfield  {journal} {\bibinfo  {journal} {Nature Photonics}\ }\textbf {\bibinfo {volume} {4}},\ \bibinfo {pages} {3} (\bibinfo {year} {2010})}\BibitemShut {NoStop}%
\bibitem [{\citenamefont {Uppu}\ \emph {et~al.}(2020)\citenamefont {Uppu}, \citenamefont {Pedersen}, \citenamefont {Wang}, \citenamefont {Olesen}, \citenamefont {Papon}, \citenamefont {Zhou}, \citenamefont {Midolo}, \citenamefont {Scholz}, \citenamefont {Wieck}, \citenamefont {Ludwig},\ and\ \citenamefont {Lodahl}}]{Uppu2020}%
  \BibitemOpen
  \bibfield  {author} {\bibinfo {author} {\bibfnamefont {R.}~\bibnamefont {Uppu}}, \bibinfo {author} {\bibfnamefont {F.~T.}\ \bibnamefont {Pedersen}}, \bibinfo {author} {\bibfnamefont {Y.}~\bibnamefont {Wang}}, \bibinfo {author} {\bibfnamefont {C.~T.}\ \bibnamefont {Olesen}}, \bibinfo {author} {\bibfnamefont {C.}~\bibnamefont {Papon}}, \bibinfo {author} {\bibfnamefont {X.}~\bibnamefont {Zhou}}, \bibinfo {author} {\bibfnamefont {L.}~\bibnamefont {Midolo}}, \bibinfo {author} {\bibfnamefont {S.}~\bibnamefont {Scholz}}, \bibinfo {author} {\bibfnamefont {A.~D.}\ \bibnamefont {Wieck}}, \bibinfo {author} {\bibfnamefont {A.}~\bibnamefont {Ludwig}},\ and\ \bibinfo {author} {\bibfnamefont {P.}~\bibnamefont {Lodahl}},\ }\bibfield  {title} {\bibinfo {title} {Scalable integrated single-photon source},\ }\href {https://doi.org/10.1126/sciadv.abc8268} {\bibfield  {journal} {\bibinfo  {journal} {Science Advances}\ }\textbf {\bibinfo {volume} {6}},\ \bibinfo {pages} {eabc8268} (\bibinfo {year} {2020})},\ \Eprint
  {https://arxiv.org/abs/https://www.science.org/doi/pdf/10.1126/sciadv.abc8268} {https://www.science.org/doi/pdf/10.1126/sciadv.abc8268} \BibitemShut {NoStop}%
\bibitem [{\citenamefont {Luo}\ \emph {et~al.}(2019)\citenamefont {Luo}, \citenamefont {Brauner}, \citenamefont {Eigner}, \citenamefont {Sharapova}, \citenamefont {Ricken}, \citenamefont {Meier}, \citenamefont {Herrmann},\ and\ \citenamefont {Silberhorn}}]{Luo2019}%
  \BibitemOpen
  \bibfield  {author} {\bibinfo {author} {\bibfnamefont {K.-H.}\ \bibnamefont {Luo}}, \bibinfo {author} {\bibfnamefont {S.}~\bibnamefont {Brauner}}, \bibinfo {author} {\bibfnamefont {C.}~\bibnamefont {Eigner}}, \bibinfo {author} {\bibfnamefont {P.~R.}\ \bibnamefont {Sharapova}}, \bibinfo {author} {\bibfnamefont {R.}~\bibnamefont {Ricken}}, \bibinfo {author} {\bibfnamefont {T.}~\bibnamefont {Meier}}, \bibinfo {author} {\bibfnamefont {H.}~\bibnamefont {Herrmann}},\ and\ \bibinfo {author} {\bibfnamefont {C.}~\bibnamefont {Silberhorn}},\ }\bibfield  {title} {\bibinfo {title} {Nonlinear integrated quantum electro-optic circuits},\ }\href {https://doi.org/10.1126/sciadv.aat1451} {\bibfield  {journal} {\bibinfo  {journal} {Science Advances}\ }\textbf {\bibinfo {volume} {5}},\ \bibinfo {pages} {eaat1451} (\bibinfo {year} {2019})},\ \Eprint {https://arxiv.org/abs/https://www.science.org/doi/pdf/10.1126/sciadv.aat1451} {https://www.science.org/doi/pdf/10.1126/sciadv.aat1451} \BibitemShut {NoStop}%
\bibitem [{\citenamefont {Pirandola}\ \emph {et~al.}(2018)\citenamefont {Pirandola}, \citenamefont {Bardhan}, \citenamefont {Gehring}, \citenamefont {Weedbrook},\ and\ \citenamefont {Lloyd}}]{Pirandola2018}%
  \BibitemOpen
  \bibfield  {author} {\bibinfo {author} {\bibfnamefont {S.}~\bibnamefont {Pirandola}}, \bibinfo {author} {\bibfnamefont {B.~R.}\ \bibnamefont {Bardhan}}, \bibinfo {author} {\bibfnamefont {T.}~\bibnamefont {Gehring}}, \bibinfo {author} {\bibfnamefont {C.}~\bibnamefont {Weedbrook}},\ and\ \bibinfo {author} {\bibfnamefont {S.}~\bibnamefont {Lloyd}},\ }\bibfield  {title} {\bibinfo {title} {Advances in photonic quantum sensing},\ }\href {https://doi.org/10.1038/s41566-018-0301-6} {\bibfield  {journal} {\bibinfo  {journal} {Nature Photonics}\ }\textbf {\bibinfo {volume} {12}},\ \bibinfo {pages} {724} (\bibinfo {year} {2018})}\BibitemShut {NoStop}%
\bibitem [{\citenamefont {Gallot}\ and\ \citenamefont {Grischkowsky}(1999)}]{Gallot1999}%
  \BibitemOpen
  \bibfield  {author} {\bibinfo {author} {\bibfnamefont {G.}~\bibnamefont {Gallot}}\ and\ \bibinfo {author} {\bibfnamefont {D.}~\bibnamefont {Grischkowsky}},\ }\bibfield  {title} {\bibinfo {title} {Electro-optic detection of terahertz radiation},\ }\href {https://doi.org/10.1364/JOSAB.16.001204} {\bibfield  {journal} {\bibinfo  {journal} {J. Opt. Soc. Am. B}\ }\textbf {\bibinfo {volume} {16}},\ \bibinfo {pages} {1204} (\bibinfo {year} {1999})}\BibitemShut {NoStop}%
\bibitem [{\citenamefont {Riek}\ \emph {et~al.}(2015)\citenamefont {Riek}, \citenamefont {Seletskiy}, \citenamefont {Moskalenko}, \citenamefont {Schmidt}, \citenamefont {Krauspe}, \citenamefont {Eckart}, \citenamefont {Eggert}, \citenamefont {Burkard},\ and\ \citenamefont {Leitenstorfer}}]{Riek2015}%
  \BibitemOpen
  \bibfield  {author} {\bibinfo {author} {\bibfnamefont {C.}~\bibnamefont {Riek}}, \bibinfo {author} {\bibfnamefont {D.~V.}\ \bibnamefont {Seletskiy}}, \bibinfo {author} {\bibfnamefont {A.~S.}\ \bibnamefont {Moskalenko}}, \bibinfo {author} {\bibfnamefont {J.~F.}\ \bibnamefont {Schmidt}}, \bibinfo {author} {\bibfnamefont {P.}~\bibnamefont {Krauspe}}, \bibinfo {author} {\bibfnamefont {S.}~\bibnamefont {Eckart}}, \bibinfo {author} {\bibfnamefont {S.}~\bibnamefont {Eggert}}, \bibinfo {author} {\bibfnamefont {G.}~\bibnamefont {Burkard}},\ and\ \bibinfo {author} {\bibfnamefont {A.}~\bibnamefont {Leitenstorfer}},\ }\bibfield  {title} {\bibinfo {title} {Direct sampling of electric-field vacuum fluctuations},\ }\href {https://doi.org/10.1126/science.aac9788} {\bibfield  {journal} {\bibinfo  {journal} {Science}\ }\textbf {\bibinfo {volume} {350}},\ \bibinfo {pages} {420} (\bibinfo {year} {2015})},\ \Eprint {https://arxiv.org/abs/https://www.science.org/doi/pdf/10.1126/science.aac9788}
  {https://www.science.org/doi/pdf/10.1126/science.aac9788} \BibitemShut {NoStop}%
\bibitem [{\citenamefont {Riek}\ \emph {et~al.}(2017)\citenamefont {Riek}, \citenamefont {Sulzer}, \citenamefont {Seeger}, \citenamefont {Moskalenko}, \citenamefont {Burkard}, \citenamefont {Seletskiy},\ and\ \citenamefont {Leitenstorfer}}]{Riek2017}%
  \BibitemOpen
  \bibfield  {author} {\bibinfo {author} {\bibfnamefont {C.}~\bibnamefont {Riek}}, \bibinfo {author} {\bibfnamefont {P.}~\bibnamefont {Sulzer}}, \bibinfo {author} {\bibfnamefont {M.}~\bibnamefont {Seeger}}, \bibinfo {author} {\bibfnamefont {A.~S.}\ \bibnamefont {Moskalenko}}, \bibinfo {author} {\bibfnamefont {G.}~\bibnamefont {Burkard}}, \bibinfo {author} {\bibfnamefont {D.~V.}\ \bibnamefont {Seletskiy}},\ and\ \bibinfo {author} {\bibfnamefont {A.}~\bibnamefont {Leitenstorfer}},\ }\bibfield  {title} {\bibinfo {title} {Subcycle quantum electrodynamics},\ }\href {https://doi.org/10.1038/nature21024} {\bibfield  {journal} {\bibinfo  {journal} {Nature}\ }\textbf {\bibinfo {volume} {541}},\ \bibinfo {pages} {376} (\bibinfo {year} {2017})}\BibitemShut {NoStop}%
\bibitem [{\citenamefont {Kizmann}\ \emph {et~al.}(2019)\citenamefont {Kizmann}, \citenamefont {Guedes}, \citenamefont {Seletskiy}, \citenamefont {Moskalenko}, \citenamefont {Leitenstorfer},\ and\ \citenamefont {Burkard}}]{Kizmann2019}%
  \BibitemOpen
  \bibfield  {author} {\bibinfo {author} {\bibfnamefont {M.}~\bibnamefont {Kizmann}}, \bibinfo {author} {\bibfnamefont {T.~L. d.~M.}\ \bibnamefont {Guedes}}, \bibinfo {author} {\bibfnamefont {D.~V.}\ \bibnamefont {Seletskiy}}, \bibinfo {author} {\bibfnamefont {A.~S.}\ \bibnamefont {Moskalenko}}, \bibinfo {author} {\bibfnamefont {A.}~\bibnamefont {Leitenstorfer}},\ and\ \bibinfo {author} {\bibfnamefont {G.}~\bibnamefont {Burkard}},\ }\bibfield  {title} {\bibinfo {title} {Subcycle squeezing of light from a time flow perspective},\ }\href {https://doi.org/10.1038/s41567-019-0560-2} {\bibfield  {journal} {\bibinfo  {journal} {Nature Physics}\ }\textbf {\bibinfo {volume} {15}},\ \bibinfo {pages} {960} (\bibinfo {year} {2019})}\BibitemShut {NoStop}%
\bibitem [{\citenamefont {Guedes}\ \emph {et~al.}(2019)\citenamefont {Guedes}, \citenamefont {Kizmann}, \citenamefont {Seletskiy}, \citenamefont {Leitenstorfer}, \citenamefont {Burkard},\ and\ \citenamefont {Moskalenko}}]{Guedes2019}%
  \BibitemOpen
  \bibfield  {author} {\bibinfo {author} {\bibfnamefont {T.~L.~M.}\ \bibnamefont {Guedes}}, \bibinfo {author} {\bibfnamefont {M.}~\bibnamefont {Kizmann}}, \bibinfo {author} {\bibfnamefont {D.~V.}\ \bibnamefont {Seletskiy}}, \bibinfo {author} {\bibfnamefont {A.}~\bibnamefont {Leitenstorfer}}, \bibinfo {author} {\bibfnamefont {G.}~\bibnamefont {Burkard}},\ and\ \bibinfo {author} {\bibfnamefont {A.~S.}\ \bibnamefont {Moskalenko}},\ }\bibfield  {title} {\bibinfo {title} {Spectra of ultrabroadband squeezed pulses and the finite-time unruh-davies effect},\ }\href {https://doi.org/10.1103/PhysRevLett.122.053604} {\bibfield  {journal} {\bibinfo  {journal} {Phys. Rev. Lett.}\ }\textbf {\bibinfo {volume} {122}},\ \bibinfo {pages} {053604} (\bibinfo {year} {2019})}\BibitemShut {NoStop}%
\bibitem [{\citenamefont {Sulzer}\ \emph {et~al.}(2020)\citenamefont {Sulzer}, \citenamefont {Oguchi}, \citenamefont {Huster}, \citenamefont {Kizmann}, \citenamefont {Guedes}, \citenamefont {Liehl}, \citenamefont {Beckh}, \citenamefont {Moskalenko}, \citenamefont {Burkard}, \citenamefont {Seletskiy},\ and\ \citenamefont {Leitenstorfer}}]{Sulzer2020}%
  \BibitemOpen
  \bibfield  {author} {\bibinfo {author} {\bibfnamefont {P.}~\bibnamefont {Sulzer}}, \bibinfo {author} {\bibfnamefont {K.}~\bibnamefont {Oguchi}}, \bibinfo {author} {\bibfnamefont {J.}~\bibnamefont {Huster}}, \bibinfo {author} {\bibfnamefont {M.}~\bibnamefont {Kizmann}}, \bibinfo {author} {\bibfnamefont {T.~L.~M.}\ \bibnamefont {Guedes}}, \bibinfo {author} {\bibfnamefont {A.}~\bibnamefont {Liehl}}, \bibinfo {author} {\bibfnamefont {C.}~\bibnamefont {Beckh}}, \bibinfo {author} {\bibfnamefont {A.~S.}\ \bibnamefont {Moskalenko}}, \bibinfo {author} {\bibfnamefont {G.}~\bibnamefont {Burkard}}, \bibinfo {author} {\bibfnamefont {D.~V.}\ \bibnamefont {Seletskiy}},\ and\ \bibinfo {author} {\bibfnamefont {A.}~\bibnamefont {Leitenstorfer}},\ }\bibfield  {title} {\bibinfo {title} {Determination of the electric field and its hilbert transform in femtosecond electro-optic sampling},\ }\href {https://doi.org/10.1103/PhysRevA.101.033821} {\bibfield  {journal} {\bibinfo  {journal} {Phys. Rev. A}\ }\textbf {\bibinfo {volume}
  {101}},\ \bibinfo {pages} {033821} (\bibinfo {year} {2020})}\BibitemShut {NoStop}%
\bibitem [{\citenamefont {Virally}\ \emph {et~al.}(2021)\citenamefont {Virally}, \citenamefont {Cusson},\ and\ \citenamefont {Seletskiy}}]{Virally2021}%
  \BibitemOpen
  \bibfield  {author} {\bibinfo {author} {\bibfnamefont {S.}~\bibnamefont {Virally}}, \bibinfo {author} {\bibfnamefont {P.}~\bibnamefont {Cusson}},\ and\ \bibinfo {author} {\bibfnamefont {D.~V.}\ \bibnamefont {Seletskiy}},\ }\bibfield  {title} {\bibinfo {title} {Enhanced electro-optic sampling with quantum probes},\ }\href {https://doi.org/10.1103/PhysRevLett.127.270504} {\bibfield  {journal} {\bibinfo  {journal} {Phys. Rev. Lett.}\ }\textbf {\bibinfo {volume} {127}},\ \bibinfo {pages} {270504} (\bibinfo {year} {2021})}\BibitemShut {NoStop}%
\bibitem [{\citenamefont {Benea-Chelmus}\ \emph {et~al.}(2016)\citenamefont {Benea-Chelmus}, \citenamefont {Bonzon}, \citenamefont {Maissen}, \citenamefont {Scalari}, \citenamefont {Beck},\ and\ \citenamefont {Faist}}]{Benea-Chelmus2016}%
  \BibitemOpen
  \bibfield  {author} {\bibinfo {author} {\bibfnamefont {I.-C.}\ \bibnamefont {Benea-Chelmus}}, \bibinfo {author} {\bibfnamefont {C.}~\bibnamefont {Bonzon}}, \bibinfo {author} {\bibfnamefont {C.}~\bibnamefont {Maissen}}, \bibinfo {author} {\bibfnamefont {G.}~\bibnamefont {Scalari}}, \bibinfo {author} {\bibfnamefont {M.}~\bibnamefont {Beck}},\ and\ \bibinfo {author} {\bibfnamefont {J.}~\bibnamefont {Faist}},\ }\bibfield  {title} {\bibinfo {title} {Subcycle measurement of intensity correlations in the terahertz frequency range},\ }\href {https://doi.org/10.1103/PhysRevA.93.043812} {\bibfield  {journal} {\bibinfo  {journal} {Phys. Rev. A}\ }\textbf {\bibinfo {volume} {93}},\ \bibinfo {pages} {043812} (\bibinfo {year} {2016})}\BibitemShut {NoStop}%
\bibitem [{\citenamefont {Benea-Chelmus}\ \emph {et~al.}(2019)\citenamefont {Benea-Chelmus}, \citenamefont {Settembrini}, \citenamefont {Scalari},\ and\ \citenamefont {Faist}}]{Benea-Chelmus2019}%
  \BibitemOpen
  \bibfield  {author} {\bibinfo {author} {\bibfnamefont {I.-C.}\ \bibnamefont {Benea-Chelmus}}, \bibinfo {author} {\bibfnamefont {F.~F.}\ \bibnamefont {Settembrini}}, \bibinfo {author} {\bibfnamefont {G.}~\bibnamefont {Scalari}},\ and\ \bibinfo {author} {\bibfnamefont {J.}~\bibnamefont {Faist}},\ }\bibfield  {title} {\bibinfo {title} {Electric field correlation measurements on the electromagnetic vacuum state},\ }\href {https://doi.org/10.1038/s41586-019-1083-9} {\bibfield  {journal} {\bibinfo  {journal} {Nature}\ }\textbf {\bibinfo {volume} {568}},\ \bibinfo {pages} {202} (\bibinfo {year} {2019})}\BibitemShut {NoStop}%
\bibitem [{\citenamefont {Settembrini}\ \emph {et~al.}(2022)\citenamefont {Settembrini}, \citenamefont {Lindel}, \citenamefont {Herter}, \citenamefont {Buhmann},\ and\ \citenamefont {Faist}}]{Settembrini2022}%
  \BibitemOpen
  \bibfield  {author} {\bibinfo {author} {\bibfnamefont {F.~F.}\ \bibnamefont {Settembrini}}, \bibinfo {author} {\bibfnamefont {F.}~\bibnamefont {Lindel}}, \bibinfo {author} {\bibfnamefont {A.~M.}\ \bibnamefont {Herter}}, \bibinfo {author} {\bibfnamefont {S.~Y.}\ \bibnamefont {Buhmann}},\ and\ \bibinfo {author} {\bibfnamefont {J.}~\bibnamefont {Faist}},\ }\bibfield  {title} {\bibinfo {title} {Detection of quantum-vacuum field correlations outside the light cone},\ }\href {https://doi.org/10.1038/s41467-022-31081-1} {\bibfield  {journal} {\bibinfo  {journal} {Nature Communications}\ }\textbf {\bibinfo {volume} {13}},\ \bibinfo {pages} {3383} (\bibinfo {year} {2022})}\BibitemShut {NoStop}%
\bibitem [{\citenamefont {Valentini}(1991)}]{Valentini1991}%
  \BibitemOpen
  \bibfield  {author} {\bibinfo {author} {\bibfnamefont {A.}~\bibnamefont {Valentini}},\ }\bibfield  {title} {\bibinfo {title} {Non-local correlations in quantum electrodynamics},\ }\href {https://doi.org/https://doi.org/10.1016/0375-9601(91)90952-5} {\bibfield  {journal} {\bibinfo  {journal} {Physics Letters A}\ }\textbf {\bibinfo {volume} {153}},\ \bibinfo {pages} {321} (\bibinfo {year} {1991})}\BibitemShut {NoStop}%
\bibitem [{\citenamefont {Guedes}\ \emph {et~al.}(2023)\citenamefont {Guedes}, \citenamefont {Vakulchyk}, \citenamefont {Seletskiy}, \citenamefont {Leitenstorfer}, \citenamefont {Moskalenko},\ and\ \citenamefont {Burkard}}]{Guedes2022}%
  \BibitemOpen
  \bibfield  {author} {\bibinfo {author} {\bibfnamefont {T.~L.~M.}\ \bibnamefont {Guedes}}, \bibinfo {author} {\bibfnamefont {I.}~\bibnamefont {Vakulchyk}}, \bibinfo {author} {\bibfnamefont {D.~V.}\ \bibnamefont {Seletskiy}}, \bibinfo {author} {\bibfnamefont {A.}~\bibnamefont {Leitenstorfer}}, \bibinfo {author} {\bibfnamefont {A.~S.}\ \bibnamefont {Moskalenko}},\ and\ \bibinfo {author} {\bibfnamefont {G.}~\bibnamefont {Burkard}},\ }\bibfield  {title} {\bibinfo {title} {Back action in quantum electro-optic sampling of electromagnetic vacuum fluctuations},\ }\href {https://doi.org/10.1103/PhysRevResearch.5.013151} {\bibfield  {journal} {\bibinfo  {journal} {Phys. Rev. Res.}\ }\textbf {\bibinfo {volume} {5}},\ \bibinfo {pages} {013151} (\bibinfo {year} {2023})}\BibitemShut {NoStop}%
\bibitem [{\citenamefont {Gündoğdu}\ \emph {et~al.}()\citenamefont {Gündoğdu}, \citenamefont {Virally}, \citenamefont {Scaglia}, \citenamefont {Seletskiy},\ and\ \citenamefont {Moskalenko}}]{Gundogdu2022}%
  \BibitemOpen
  \bibfield  {author} {\bibinfo {author} {\bibfnamefont {S.}~\bibnamefont {Gündoğdu}}, \bibinfo {author} {\bibfnamefont {S.}~\bibnamefont {Virally}}, \bibinfo {author} {\bibfnamefont {M.}~\bibnamefont {Scaglia}}, \bibinfo {author} {\bibfnamefont {D.~V.}\ \bibnamefont {Seletskiy}},\ and\ \bibinfo {author} {\bibfnamefont {A.~S.}\ \bibnamefont {Moskalenko}},\ }\bibfield  {title} {\bibinfo {title} {Self-referenced subcycle metrology of quantum fields},\ }\href {https://doi.org/https://doi.org/10.1002/lpor.202200706} {\bibfield  {journal} {\bibinfo  {journal} {Laser \& Photonics Reviews}\ }\textbf {\bibinfo {volume} {n/a}},\ \bibinfo {pages} {2200706}},\ \Eprint {https://arxiv.org/abs/https://onlinelibrary.wiley.com/doi/pdf/10.1002/lpor.202200706} {https://onlinelibrary.wiley.com/doi/pdf/10.1002/lpor.202200706} \BibitemShut {NoStop}%
\bibitem [{\citenamefont {van~der Valk}\ \emph {et~al.}(2004)\citenamefont {van~der Valk}, \citenamefont {Wenckebach},\ and\ \citenamefont {Planken}}]{vanderValk2004}%
  \BibitemOpen
  \bibfield  {author} {\bibinfo {author} {\bibfnamefont {N.~C.~J.}\ \bibnamefont {van~der Valk}}, \bibinfo {author} {\bibfnamefont {T.}~\bibnamefont {Wenckebach}},\ and\ \bibinfo {author} {\bibfnamefont {P.~C.~M.}\ \bibnamefont {Planken}},\ }\bibfield  {title} {\bibinfo {title} {Full mathematical description of electro-optic detection in optically isotropic crystals},\ }\href {https://doi.org/10.1364/JOSAB.21.000622} {\bibfield  {journal} {\bibinfo  {journal} {J. Opt. Soc. Am. B}\ }\textbf {\bibinfo {volume} {21}},\ \bibinfo {pages} {622} (\bibinfo {year} {2004})}\BibitemShut {NoStop}%
\bibitem [{\citenamefont {Planken}\ \emph {et~al.}(2001)\citenamefont {Planken}, \citenamefont {Nienhuys}, \citenamefont {Bakker},\ and\ \citenamefont {Wenckebach}}]{Planken2001}%
  \BibitemOpen
  \bibfield  {author} {\bibinfo {author} {\bibfnamefont {P.~C.~M.}\ \bibnamefont {Planken}}, \bibinfo {author} {\bibfnamefont {H.-K.}\ \bibnamefont {Nienhuys}}, \bibinfo {author} {\bibfnamefont {H.~J.}\ \bibnamefont {Bakker}},\ and\ \bibinfo {author} {\bibfnamefont {T.}~\bibnamefont {Wenckebach}},\ }\bibfield  {title} {\bibinfo {title} {Measurement and calculation of the orientation dependence of terahertz pulse detection in znte},\ }\href {https://doi.org/10.1364/JOSAB.18.000313} {\bibfield  {journal} {\bibinfo  {journal} {J. Opt. Soc. Am. B}\ }\textbf {\bibinfo {volume} {18}},\ \bibinfo {pages} {313} (\bibinfo {year} {2001})}\BibitemShut {NoStop}%
\bibitem [{\citenamefont {Chen}\ \emph {et~al.}(2009)\citenamefont {Chen}, \citenamefont {He}, \citenamefont {Shen}, \citenamefont {Li~Zhao}, \citenamefont {Xu}, \citenamefont {Wang}, \citenamefont {Wang},\ and\ \citenamefont {Dai}}]{Chen2009}%
  \BibitemOpen
  \bibfield  {author} {\bibinfo {author} {\bibfnamefont {X.}~\bibnamefont {Chen}}, \bibinfo {author} {\bibfnamefont {S.}~\bibnamefont {He}}, \bibinfo {author} {\bibfnamefont {Z.}~\bibnamefont {Shen}}, \bibinfo {author} {\bibfnamefont {F.}~\bibnamefont {Li~Zhao}}, \bibinfo {author} {\bibfnamefont {K.~Y.}\ \bibnamefont {Xu}}, \bibinfo {author} {\bibfnamefont {G.}~\bibnamefont {Wang}}, \bibinfo {author} {\bibfnamefont {R.}~\bibnamefont {Wang}},\ and\ \bibinfo {author} {\bibfnamefont {N.}~\bibnamefont {Dai}},\ }\bibfield  {title} {\bibinfo {title} {Influence of nonlinear effects in znte on generation and detection of terahertz waves},\ }\href {https://doi.org/10.1063/1.3068480} {\bibfield  {journal} {\bibinfo  {journal} {Journal of Applied Physics}\ }\textbf {\bibinfo {volume} {105}},\ \bibinfo {pages} {023106} (\bibinfo {year} {2009})},\ \Eprint {https://arxiv.org/abs/https://doi.org/10.1063/1.3068480} {https://doi.org/10.1063/1.3068480} \BibitemShut {NoStop}%
\bibitem [{\citenamefont {Caumes}\ \emph {et~al.}(2002)\citenamefont {Caumes}, \citenamefont {Videau}, \citenamefont {Rouyer},\ and\ \citenamefont {Freysz}}]{Caumes2002}%
  \BibitemOpen
  \bibfield  {author} {\bibinfo {author} {\bibfnamefont {J.-P.}\ \bibnamefont {Caumes}}, \bibinfo {author} {\bibfnamefont {L.}~\bibnamefont {Videau}}, \bibinfo {author} {\bibfnamefont {C.}~\bibnamefont {Rouyer}},\ and\ \bibinfo {author} {\bibfnamefont {E.}~\bibnamefont {Freysz}},\ }\bibfield  {title} {\bibinfo {title} {Kerr-like nonlinearity induced via terahertz generation and the electro-optical effect in zinc blende crystals},\ }\href {https://doi.org/10.1103/PhysRevLett.89.047401} {\bibfield  {journal} {\bibinfo  {journal} {Phys. Rev. Lett.}\ }\textbf {\bibinfo {volume} {89}},\ \bibinfo {pages} {047401} (\bibinfo {year} {2002})}\BibitemShut {NoStop}%
\bibitem [{\citenamefont {Tian}\ \emph {et~al.}(2008)\citenamefont {Tian}, \citenamefont {Wang}, \citenamefont {Xing}, \citenamefont {Gu}, \citenamefont {Li}, \citenamefont {He}, \citenamefont {Chai}, \citenamefont {Wang},\ and\ \citenamefont {Zhang}}]{Zhen2008}%
  \BibitemOpen
  \bibfield  {author} {\bibinfo {author} {\bibfnamefont {Z.}~\bibnamefont {Tian}}, \bibinfo {author} {\bibfnamefont {C.}~\bibnamefont {Wang}}, \bibinfo {author} {\bibfnamefont {Q.}~\bibnamefont {Xing}}, \bibinfo {author} {\bibfnamefont {J.}~\bibnamefont {Gu}}, \bibinfo {author} {\bibfnamefont {Y.}~\bibnamefont {Li}}, \bibinfo {author} {\bibfnamefont {M.}~\bibnamefont {He}}, \bibinfo {author} {\bibfnamefont {L.}~\bibnamefont {Chai}}, \bibinfo {author} {\bibfnamefont {Q.}~\bibnamefont {Wang}},\ and\ \bibinfo {author} {\bibfnamefont {W.}~\bibnamefont {Zhang}},\ }\bibfield  {title} {\bibinfo {title} {Quantitative analysis of kerr nonlinearity and kerr-like nonlinearity induced via terahertz generation in znte},\ }\href {https://doi.org/10.1063/1.2838446} {\bibfield  {journal} {\bibinfo  {journal} {Applied Physics Letters}\ }\textbf {\bibinfo {volume} {92}},\ \bibinfo {pages} {041106} (\bibinfo {year} {2008})},\ \Eprint {https://arxiv.org/abs/https://doi.org/10.1063/1.2838446} {https://doi.org/10.1063/1.2838446}
  \BibitemShut {NoStop}%
\bibitem [{\citenamefont {Moskalenko}\ \emph {et~al.}(2015)\citenamefont {Moskalenko}, \citenamefont {Riek}, \citenamefont {Seletskiy}, \citenamefont {Burkard},\ and\ \citenamefont {Leitenstorfer}}]{Moskalenko2015}%
  \BibitemOpen
  \bibfield  {author} {\bibinfo {author} {\bibfnamefont {A.~S.}\ \bibnamefont {Moskalenko}}, \bibinfo {author} {\bibfnamefont {C.}~\bibnamefont {Riek}}, \bibinfo {author} {\bibfnamefont {D.~V.}\ \bibnamefont {Seletskiy}}, \bibinfo {author} {\bibfnamefont {G.}~\bibnamefont {Burkard}},\ and\ \bibinfo {author} {\bibfnamefont {A.}~\bibnamefont {Leitenstorfer}},\ }\bibfield  {title} {\bibinfo {title} {Paraxial theory of direct electro-optic sampling of the quantum vacuum},\ }\href {https://doi.org/10.1103/PhysRevLett.115.263601} {\bibfield  {journal} {\bibinfo  {journal} {Phys. Rev. Lett.}\ }\textbf {\bibinfo {volume} {115}},\ \bibinfo {pages} {263601} (\bibinfo {year} {2015})}\BibitemShut {NoStop}%
\bibitem [{\citenamefont {Siegman}(1986)}]{siegman1986}%
  \BibitemOpen
  \bibfield  {author} {\bibinfo {author} {\bibfnamefont {A.~E.}\ \bibnamefont {Siegman}},\ }\href@noop {} {\emph {\bibinfo {title} {Lasers}}}\ (\bibinfo  {publisher} {University science books},\ \bibinfo {year} {1986})\BibitemShut {NoStop}%
\end{thebibliography}%

\end{document}